\documentclass[amsthm,final]{autart}

\usepackage{graphicx}
\usepackage{epstopdf}

\usepackage{amsmath}
\usepackage{lineno}
\usepackage{multirow}
\usepackage{amsfonts}
\usepackage{textcomp}
\usepackage{stfloats}

\usepackage{amssymb}
\usepackage{natbib}
\usepackage{xcolor}
\usepackage[bookmarksnumbered, colorlinks, citecolor=blue!50!black!100, linkcolor=blue, urlcolor=black]{hyperref}
\usepackage{float}
\usepackage{indentfirst}
\usepackage{mathrsfs}

\usepackage{stmaryrd}

\theoremstyle{plain}
\newtheorem{remark}{Remark}

\newtheorem{definition}{Definition}
\newtheorem{theorem}{Theorem}
\newtheorem{lemma}{Lemma}
\newtheorem{problem}{Problem}
\newtheorem{proposition}{Proposition}
\newtheorem{corollary}{Corollary}

\begin{document}

\begin{frontmatter}

\title{ Recursive Privacy-Preserving Estimation  \\ Over Markov Fading Channels \thanksref{footnoteinfo}} % Title, preferably not more 
                                                % than 10 words.

\thanks[footnoteinfo]{This work was supported by the National Natural Science Foundation of China under grant 61733009, National Key Research and Development Program of China under grant 2022YFB25031103, and Huaneng Group Science and Technology Research Project under Grant HNKJ22-H105. 
\textit{(Corresponding author: Xiao He.)}
}

\author[Macao]{Jie Huang}\ead{huangjie18@ tsinghua.org.cn},    
\author[Edmonton]{Fanlin Jia}\ead{fanlin@ualberta.ca},               
\author[Beijing]{Xiao He}\ead{hexiao@tsinghua.edu.cn}  

\address[Macao]{Faculty of Science and Technology, University of Macau, Taipa, Macao, P.~R.~China}   
\address[Edmonton]{Department of Electrical and Computer Engineering, University of Alberta, Edmonton AB T6G 1H9, Canada}
\address[Beijing]{Department of Automation, Tsinghua University, Beijing 100084, P.~R.~China}  
                                         

\begin{abstract} 
	In industrial applications, the presence of moving machinery, vehicles, and personnel, contributes to the dynamic nature of the wireless channel. This time variability induces channel fading, which can be effectively modeled using a Markov fading channel (MFC). 
	In this paper, we investigate the problem of secure state estimation for systems that communicate over a MFC in the presence of an eavesdropper. The objective is to enable a remote authorized user to accurately estimate the states of a dynamic system, while considering the potential interception of the sensor's packet through a wiretap channel. To prevent information leakage, a novel co-design strategy is established, which combines a privacy-preserving mechanism with a state estimator. To implement our encoding scheme, a nonlinear mapping of the innovation is introduced based on the weighted reconstructed innovation previously received by the legitimate user. Corresponding to this encoding scheme, we design a recursive privacy-preserving filtering algorithm to achieve accurate estimation. The boundedness of estimation error dynamics at the legitimate user's side is discussed and the divergence of the eavesdropper's estimation error is analyzed, which demonstrates the effectiveness of our co-design strategy in ensuring secrecy. Furthermore, a simulation example of a three-tank system is provided to demonstrate the effectiveness and feasibility of our privacy-preserving estimation method.
\end{abstract}

\end{frontmatter}

\small

\section{Introduction}

	Cyber-physical systems facilitate the transformation of traditional industries into intelligent manufacturing through real-time interaction and collaboration among systems and devices across different regions. The integration of communication technologies not only enhances capacities, efficiency and quality but also entails risks in industrial data security. Eavesdropping is indeed a significant data security challenge faced by industrial systems~\citep{Lu2019_A,ShangTAC2023linear,crimson2023remote,ZHU2025112091,ZHOU2022110151}. Hostile eavesdroppers can exploit the broadcast nature of wireless networks to easily steal sensitive information such as measurements, references, control inputs and parameters of plant models as well as controllers ~\citep{Kogiso2015_C,AN2025112275,CRIMSON2025111932,CHEN2023110908,Umsonst2021_O}. In order to address such a threat to data confidentiality in control systems, researchers are exploring various strategies, including homomorphic encryption in \cite{Alexandru2020_C,Farokhi2017_S,Lu2018_P,Gao2023_D}, differential privacy techniques in \cite{Wang2022_D,He2020_D,Liu2020_D}, encoding-based state estimation and control in \cite{Gao2022_C,Gao2021_E,Cintuglu2019_S,Liang2023_S}. 

	Encoding-based methods offer significant advantages in computational efficiency, particularly for energy-constrained or time-critical systems, while maintaining relatively high levels of secrecy.
	Among research on developing encoding mechanisms in the physical layer of wireless communications for control systems, a class of control-theoretic approaches was proposed in \cite{Tsiamis2017_P} and \cite{Leong2017_O}. 
	By utilizing the expected error covariance of both the legitimate user and the eavesdropper to describe data availability and confidentiality, the design of a privacy-preserving mechanism (PPM) can be seamlessly integrated into the control tasks. 
	In the context of unstable systems, based on a packet-based paradigm in \cite{Sinopoli2004_K}, PPMs that withhold data were proposed in \cite{Tsiamis2017_P} and  \cite{Leong2017_S} to meet secrecy. Additionally, an event-triggered transmission strategy was designed in \cite{Lu2020_O} to guarantee secrecy without the limitation on system stability. 
	To mitigate the loss of data availability caused by withholding mechanisms, a novel class of encoding was presented in \cite {Tsiamis2017_S} and \cite {Tsiamis2018_S} for unstable and stable systems, respectively. 
	Within this encoding framework, perfect secrecy is achieved by introducing a critical event, where the legitimate user successfully receives data while the eavesdropper fails to intercept the transmission.
	Furthermore, an expectational lossless PPM was proposed in \cite {Huang2022_P} based on the weighted-reference structure to reduce computational costs without increasing the private key's data length.

	Due to the effect caused by obstacles in wave propagation, multi-path propagation and shadowing, channel states change over time. Wireless channels often experience fading, resulting in packet dropout and degradation of estimation performance and control stability. 
	To capture the temporal correlations of channel states, a Markov process is employed to construct a fading channel model, named as Markov fading channels (MFC). 
	In this paper, we adopt the MFC model proposed in \cite{Quevedo2013_S} and \cite{Liu_TAC2022_R},  which encompasses independent and identically distributed transmissions, the Gilbert-Elliot Model, and  the finite-state Markov channel.
	In \cite{Impicciatore_CDC2022_S}, a data-withholding mechanism is developed to achieve secrecy over finite-state Markov channels. However, this secrecy mechanism is specifically designed for unstable systems and relies on the condition that the eavesdropper's interception probability is lower than the legitimate user's reception probability. Additionally, the data-withholding mechanism may exacerbate issues related to data availability.

	In this paper, we explore the problem of privacy-preserving state estimation (PPSE) for systems in the presence of an eavesdropper over MFCs. The key contributions of this work are summarized as follows:
	
	\begin{enumerate}
		\item A novel co-design strategy for a PPM and a state estimator is proposed under an MFC. 
		The proposed PPM is proven to be lossless in expectation, ensuring relatively high data availability compared to the results in \cite{Impicciatore_CDC2022_S}.
		Moreover, the security guarantees do not depend on the computational and communicational capabilities of the eavesdropper.
		An recursive privacy-preserving filter (PPF) is designed to complement the PPM, achieving accurate estimation with low computational complexity.
		
		\item A lightweight PPM is introduced to enable fast computation and secure transmission.
		By incorporating a scalar weight and stochastic quantization, the proposed PPM is applicable to both unstable and stable systems while maintaining a lower computational burden, compared to matrix power operations in \cite{Tsiamis2020_S} and \cite{Tsiamis2018_S}.
		Additionally, the scalar encoding parameter can amplify the eavesdropper’s estimation error, leading to faster divergence than \cite{Huang2022_P}.
		
		\item The exponential boundedness of the legitimate user’s estimation error covariance and the divergence of the eavesdropper’s estimation error expectation are analyzed.
		The exponential boundedness analysis considers the correlation between the encoding error and the measurement noise, illustrating the impact of the PPM’s distortion rate.
		The divergence conditions are examined under the worst-case scenario where the eavesdropper intercepts all data except at the moment when the critical event occurs. The analysis is extended to show that divergence generally holds in the broader case.
		These analyses demonstrate that the co-design strategy maintains data availability and confidentiality.	
	\end{enumerate}

The remainder of this paper is organized as follows. 
%Section \uppercase\expandafter{\romannumeral2} presents a model of a remote estimation system together with a PPM and a MFC. 
%Section \uppercase\expandafter{\romannumeral3} gives the co-design of the PPM and the filter and  Section \uppercase\expandafter{\romannumeral4} discusses the performance analysis of boundedness for the legitimate user's estimation, as well as the analysis of secrecy. In Section \uppercase\expandafter{\romannumeral5}, the effectiveness of the proposed privacy-preserving filtering algorithms is evaluated numerically. Finally, Section \uppercase\expandafter{\romannumeral6} provides the conclusion.
Section 2 presents a model of a remote estimation system together with a PPM and a MFC. Section 3 gives the co-design of the PPM and the filter and  Section 4 discusses the performance analysis of boundedness for the legitimate user's estimation, as well as the analysis of secrecy. In Section 5, the effectiveness of the proposed privacy-preserving filtering algorithms is evaluated numerically. Finally, Section 6 provides the conclusion.

\textit{Notations:} $\mathbb{N}_0$ is the set of nonnegative integers.
$\mathbb{R}^{d_x}$ denotes the $d_x$-dimensional Euclidean Euclidean space. 
$\mathcal{N} \left(\mu, \Sigma \right) $ is the normal distribution with mean $\mu$ and covariance $\Sigma$.
$\mathrm{Prob} \left\{ \mathcal{E} \right\} $  denotes the probability of the event $\mathcal{E}$.
$\mathbb{E} \left[ X \right] $ and $\mathrm{Cov} \left[ X \right] $ denote the expectation and covariance of a random variable $X$.
$\left\{ \gamma_{k} \right\}_{\mathbb{N}_0}$ is the sequence $\left\{\gamma_0, \gamma_1, ... \right\}$.
$Y \le X$ denotes that the matrix $\left( X-Y \right) $ is semi-positive semidefinite.
For $X \in \mathbb{R}^{m\times n}$, $X^{\top}$ is the transpose of $X$.
For $Y \in \mathbb{R}^{m\times m}$, $\mathrm{tr}{(Y)}$ stands for the trace of $Y$.
The maximum and minimum eigenvalues of a matrix are denoted by $\lambda _{\max}\left( \cdot \right) $ and $\lambda _{\min}\left( \cdot \right) $, respectively.
$\mathrm{diag} \left\{ y \left( 1\right),y \left( 2 \right), \dots \right\}$ denotes the diagonal matrix with the diagonal elements $ y \left( 1\right),y \left( 2 \right), \dots $; $I_{n}$ is the $n$-dimensional identity matrix.
$\left\| Y \right\| $ denotes the Euclidean norm of a vector or a matrix $Y$.

\section{Problem Statement}

\subsection{Dynamic System Model }
\begin{figure}
\begin{center}
\includegraphics[height=4cm]{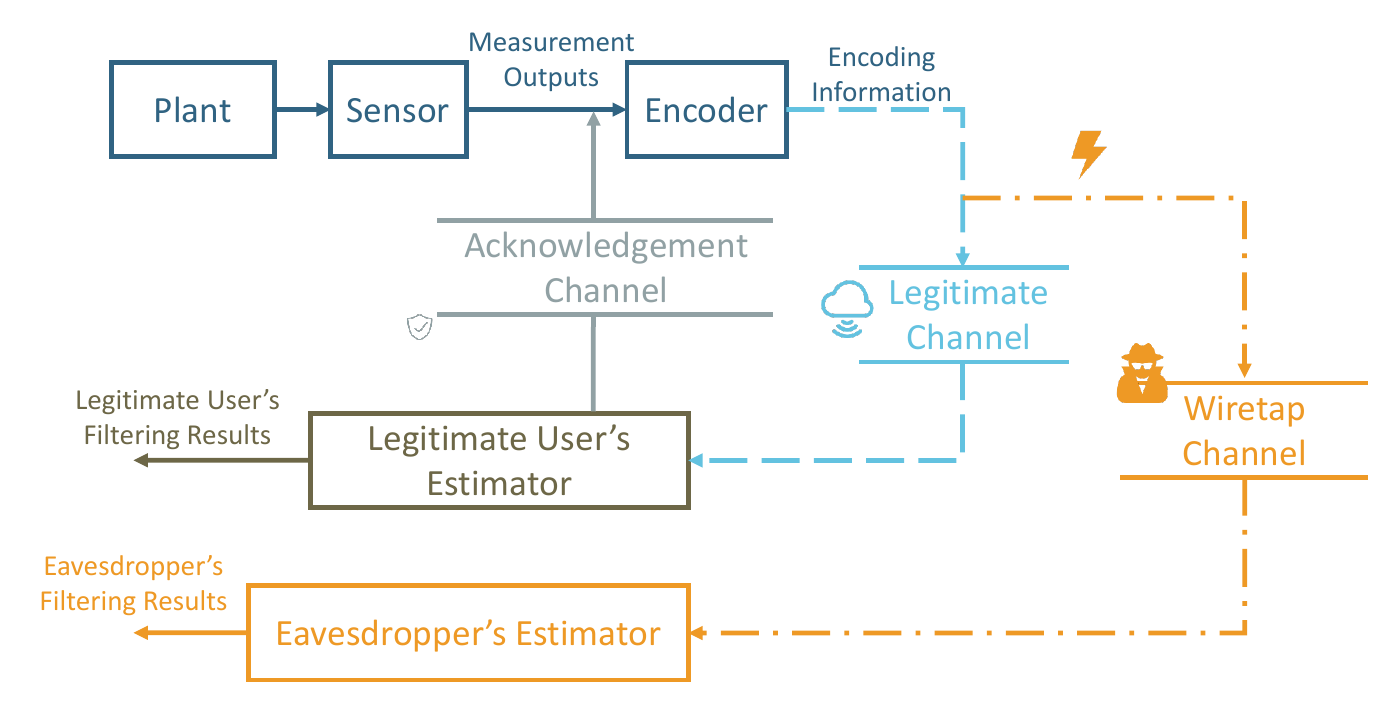}    
\caption{An architecture of remote estimation with an eavesdropper.}  
\label{Fig1}                                 
\end{center}                                 
\end{figure}

The architecture of remote estimation is depicted in Fig. \ref{Fig1}, wherein the outputs of a plant are collected by smart sensors and the innovation generated by the smart sensor are transmitted to a remote estimator via a wireless channel. Unfortunately, the wireless channel is not reliable enough so the information may be intercepted by an eavesdropper. To ensure the privacy of information, an encoding module is employed by the smart sensor to encrypt the data. The remote legitimate user realizes state estimation by using the received encoded information. 

The dynamic model of the plant is a linear discrete-time system with the following form:
\begin{equation}\label{eq:1}
	\begin{cases}	x_{k+1}=Ax_k+w_{k+1},\\	y_k ~~~ =Cx_k+v_k,\\
	\end{cases}
\end{equation}
	where $x_k\in \mathbb{R} ^{d_x}$ is a state vector with $x_0 \sim \mathcal{N} \left( \bar{x}_0,\bar{P}_0 \right) $; $y_k\in \mathbb{R} ^{d_y}$is an output of the plant. The noises $w_k\in \mathbb{R} ^{d_x}$ and $v_k\in \mathbb{R} ^{d_y}$ are mutually independent, where $w_k\sim \mathcal{N} \left( 0,Q \right) $ and $v_k \sim \mathcal{N} \left( 0,R \right) $ with $Q \ge 0$ and $R > 0$. The noise signals are also independent of the initial state. It is assumed that $( A,C ) $ is observable and $( A,Q^{\frac{1}{2}} ) $ is controllable. The system and noise parameters $A$, $C$, $Q$, $R$ and $\bar{P}_0$ are all available to the sensor, the legitimate user and the eavesdropper. 

If the data are transmitted without adequate protection, there is a significant risk to the privacy of the information, since it can be easily recovered and potentially exploited by malicious users. Consequently, it is crucial to strike a balance between the estimation accuracy of the legitimate user and the estimation capability of the eavesdropper.

	\subsection{Wireless Channel Model}

In scenarios where transceivers or the wireless environment are mobile, communication between the sensor and the user occurs over a wireless fading channel, resulting in stochastic data loss. 
The channel quality fluctuates over time and exhibits temporal correlations, which can be modeled as a time-varying random process represented by a Markov chain (MC) \citep{Sadeghi_SPM2008_F,Quevedo2013_S,Liu_TAC2022_R}. Specifically, a time-homogeneous MC with $M$ states is utilized to capture shadow fading, i.e.,		
\begin{equation*}
	\rho _k\in \Theta \triangleq \left\{ \theta _1,...,\theta _M \right\}.
\end{equation*}	
The transition probability from state $\theta_n$ at time slot $k-1$ to state $\theta_m$ at time slot $k$ is given by
\begin{equation}\label{eq:2}
	p_{nm}\triangleq \mathrm{Prob}\left\{ \rho _k=\theta _m|\rho _{k-1}=\theta _n \right\},  m,n\in \left\{ 1,2,...,M \right\}.
\end{equation}

All the channel states are assumed to be aperiodic and positive recurrent, and thus the MC is ergodic with steady state probability distribution $ \pi_i= \lim_{k\to \infty} \mathrm{Prob} \left\{\rho_k=\theta_i\right\}$, for $ \theta_i \in \Theta$.

Channel fading frequently results in random packet dropouts, thereby binary stochastic processes $\left\{ \gamma _k \right\} _{\mathbb{N} _0}$ and $\left\{ \gamma _{k}^{e} \right\} _{\mathbb{N} _0}$ are introduced to model this phenomenon of the legitimate and wiretap channels. We assume that each data packet transmitted through either the legitimate or wiretap channel is received perfectly or completely lost.
Consequently, the outputs of the channels are regulated as
\begin{equation}\label{eq3}
	\zeta _k=\begin{cases}	z_k, ~if~\gamma _{k}=1,\\	0, ~if~~\gamma _{k}=0,\\\end{cases}
	\zeta_{k}^{e}=\begin{cases}	z_k, ~if~\gamma _{k}^{e}=1,\\	0, ~if~~\gamma _{k}^{e}=0,\\\end{cases}
\end{equation}
where $z_k$ represents the encoded data to be defined later.

The distributions of $\left\{ \gamma _k \right\} _{\mathbb{N} _0}$ and $\left\{ \gamma _{k}^{e} \right\} _{\mathbb{N} _0}$ are time-homogeneous and the probabilities of packet dropout are denoted by 
\begin{align}\label{eq4}
	\nonumber \mu_{k,m}&=\mathrm{Prob} \left\{ \gamma _k=0|\rho _k=\theta _m \right\} , \\ \mu_{k,m}^{e}& =\mathrm{Prob} \left\{ \gamma _{k}^{e}=0|\rho _{k}^{e}=\theta _m \right\},
\end{align}
where $\rho _{k}^{e}$ denotes the state of the wiretap channel.
When conditioned on the channel state, the transmission dropout processes are temporarily independent.
The channel state information is available at the remote estimator. The reception probability may be obtained via some standard channel estimation techniques. 
In the rest of this paper, we denote the channel outcome $\gamma_{k}$ and the output $\zeta_{k}$ conditioned on the channel state $\rho_{k}=\theta_m$, as $\gamma_{k,m}$ and $\zeta_{k,m}$, respectively.
Similarly, the outcome and output of the wiretap channel are denoted as $\gamma_{k,m}^e$ and $\zeta_{k,m}^e$.

\subsection{Encoding Scheme}

The plant outputs $y_k$ are collected and processed by smart sensors. Due to bandwidth constraints, innovation signals $\varepsilon _k\in \mathbb{R} ^{d_y}$ are transmitted to the remote estimator, instead of the plant outputs $y_k$. The innovation $\varepsilon _k $ are generated by $\varepsilon _k=y_k-CA\hat{x}_{k-1|k-1,m}$, where the state estimation $\hat{x}_{k-1|k-1,m}$ is obtained from the local estimator embedded in the smart sensor. The sensor runs the same estimator locally as the remote estimator. Moreover, it can receive the channel outcomes $\gamma _{k,m}$ via a reliable acknowledgment channel. Consequently, the sensor utilizes the channel outcomes $\gamma _{k,m}$ to determine whether to update the local estimation at a given time slot.

To prevent information leakage, the innovation signals are encoded before transmission. The encoder is defined as follows:
\begin{equation}\label{eq:8}
	z_k=f_k\left( \varepsilon _k,z_t,\gamma _{t,m},t<k \right),
\end{equation}
where $f_k\left( \cdot \right) :\mathbb{R} ^{d_y\left( k+1 \right) +d_zk}\times \left\{ 0,1 \right\} ^k\rightarrow \mathbb{R} ^{d_z}$ is a mapping function, with the dimension $d_z$ determined by the design of the encoder. 
The term $z_k$ denotes the encoded result. The encoded data $z_t$ and the channel outcome $\gamma_{t,m}$ from the previous time slot $t$ (for some $t<k$) are employed to construct the reference for encoding. We define the decoding function as $ \bar{\varepsilon } _k = \bar{f} \left( z_k, z_t, \gamma_t , t<k \right) $. 
In this paper, we utilize the decoded innovation $\bar{\varepsilon } _k $ as the reference for future moments. The functions and parameters of the encoder and decoder are accessible to the sensor, allowing the sensor to reconstruct the decoded result $\bar{\varepsilon } _k$ and use it as the reference. Based on this mechanism, the smart sensor integrates an estimator, an encoder and a decoder. Thus, the computational complexity of the encoding-decoding process and state estimation should be sufficiently low to allow for integration within the sensor.

	\subsection{Decoding State Estimator}
Due to fading introduced by the MFC, the legitimate user successfully receives the encoded data $ z_k $ only when the channel outcome $ \gamma_{k,m} = 1$. The decoding mechanism operates exclusively upon successful transmission of the encoded data, yielding decoded results denoted by $ \bar{\varepsilon}_k = \mathbb{E} \left[ \varepsilon_k | \gamma_{k,m}=1, z_k  \right] $. 
To ensure accurate recovery of the encoded information, the PPM must satisfy the following conditions:
\begin{align}\label{eq8-1}
	& \mathbb{E} \left[ \bar{\varepsilon}_k - \varepsilon _k | \gamma_{k,m}=1, z_k \right]  = 0,  \nonumber \\
	& \mathbb{E} \left[ (\bar{\varepsilon}_k - \varepsilon _k  ) (\bar{\varepsilon}_k - \varepsilon _k  )^ \top | \gamma_{k,m}=1, z_k \right] < \infty . \nonumber
\end{align}
Both the legitimate user and the eavesdropper are aware of the encoding and decoding schemes, and each estimates the system state using all information received and decoded up to time $k$.
We denote the legitimate user's information by
\begin{equation} 
	\mathcal{I}_{k,m} = \left\{ 
	\boldsymbol{\zeta}_{0:k-1}, 
	\boldsymbol{\gamma}_{0:k-1}, 
	\zeta_{k,m}, 
	\gamma_{k,m} 
	\right\}, \nonumber 
\end{equation}
and the eavesdropper's information by 
\begin{equation} 
	\mathcal{I}_{k,m}^{e} = \left\{ 
	\boldsymbol{\zeta}_{0:k-1}^{e}, 
	\boldsymbol{\gamma}_{0:k-1}^{e}, 
	\zeta_{k,m}^{e}, 
	\gamma_{k,m}^{e},
	\gamma_{0:k,m}, 
	\right\}, \nonumber
\end{equation}
where
\begin{equation} 
	\boldsymbol{\zeta}_{0:k-1}=( \zeta _{0,\bar{n}}, \dots,\zeta _{k-1,\bar{m}} ),~ \boldsymbol{\gamma}_{0:k-1}=( \gamma _{0,\bar{n}},...,\gamma _{k-1,\bar{m}} ) \nonumber
\end{equation}
are the legitimate user's received outputs and the authorized-channel outcomes up to $k-1$, for channel states $ \rho_0= \theta_{\bar{n}}, \dots, \rho_{k-1}= \theta_{\bar{m}} \in \Theta $.
Similarly,
\begin{equation} 
	\boldsymbol{\zeta }_{0:k-1}^{e}=( \zeta _{0,\breve{n}}^{e},...,\zeta _{k-1,\breve{m}}^{e} ),
	~ \boldsymbol{\gamma} _{0:k-1}^{e}=( \gamma _{0,\breve{n}}^{e},...,\gamma _{k-1,\breve{m}}^{e} ) \nonumber
\end{equation} 
are the eavesdropper's intercepted outputs and the wiretap-channel outcomes up to $k-1$, for channel states $( \rho_0 ^e = \theta_{\breve{n}}, \dots, \rho_{k-1} ^e= \theta_{\breve{m}}) \in \Theta$.
Notably, the eavesdropper also knows the legitimate user's reception history $\boldsymbol{\gamma}_{0:k}$.

With the definitions of received information, if the authorized-channel states are $ \rho_{k} = \theta_m $ and $ \rho_{k-1} = \theta_n $,the recursive mode-dependent state estimator adopted in this paper is structured as follows:
\begin{equation}\label{eq9}
	\left\{
	\begin{aligned}
		& \hat{x}_{k|k-1,n}	= \mathbb{E} \left[ x_k | \mathcal{I} _{k-1,n} \right] \\
		& \hat{x}_{k|k,m}~~~ = \mathbb{E} \left[ x_k | \mathcal{I} _{k,m} \right] \\
		& P_{k|k-1,n} = \mathrm{Cov} \left[ x_k | \mathcal{I} _{k-1,n} \right] \\		
		& P_{k|k,m}	~~~  =\mathrm{Cov} \left[ x_k|\mathcal{I} _{k,m} \right],
	\end{aligned}
	\right.
\end{equation}
where $\hat{x}_{k|k-1,n}$ represents the one-step prediction of state $x_k$ conditioned on the received information up to $k-1$ for the channel state $ \rho_{k-1} = \theta_n $, while $\hat{x}_{k|k,m}$ is the state estimate conditioned on the received information up to $k$ for the channel state $ \rho_{k} = \theta_m $. The terms $P_{k|k-1,m}$ and $P_{k|k,m}$ denote the corresponding error covariance matrices. 
The eavesdropper's estimator is defined similarly:
\begin{equation}\label{eq10}
	\left\{
	\begin{aligned}		
		&\hat{x}_{k|k-1,n}^{e}	=\mathbb{E} \left[ x_k|\mathcal{I} _{k-1,n}^{e} \right]\\
		&\hat{x}_{k|k,m}^{e}	~~~ 	 =\mathbb{E} \left[ x_k|\mathcal{I} _{k,m}^{e} \right]\\	
		&P_{k|k-1,n}^{e}		=\mathrm{Cov} \left[ x_k|\mathcal{I}_{k-1,n}^{e} \right]\\
		&P_{k|k,m}^{e}~~~ =\mathrm{Cov} \left[x_k|\mathcal{I} _{k,m}^{e} \right].
	\end{aligned}
	\right.
\end{equation}
\begin{remark}
	In the design of the PPF, we assume that the predicted state $ \hat{x}_{k|k-1,n} $ can be approximated by Gaussian distribution. 
	The assumption is a standard simplification in nonlinear filtering (see, e.g., \cite{Kotecha_TSP2003_G,Ito_TAC2000_G,Ribeiro_TSP2006_S,You_IFAC2008_M,You_ACC2009_R}). Due to the nonlinearity introduced by the PPM, the posterior density function of the state does not hold a closed-form over the time evolution. By adopting this Gaussian approximation, the filtering algorithm can be computed recursively, akin to the Kalman filter (KF), requiring only simple algebraic operations per iteration and significantly reducing computational complexity.
\end{remark}

		\subsection{Problem}

The objective of this paper is to design a PPM with the corresponding filter to achieve the definition of secrecy, as outlined below:
\begin{definition}\label{definition1}
	Given a system (\ref{eq:1}) and a channel model (\ref{eq:2})-(\ref{eq4}), a PPM  (\ref{eq:8}) is perceived to achieve secrecy if and only if both the following conditions hold:
	\begin{itemize}
		\item[(i)] The legitimate user's estimation is unbiased in terms of expectation: 
		\begin{equation}\label{eq:9}
			x_k-\mathbb{E} \left[ \hat{x}_{k|k,m} \right] =0, ~\text{when}~ \gamma_{k,m}=1.
		\end{equation}
		In addition, the legitimate user's estimation error covariance matrix should be bounded, i.e. $\exists \bar{\rho}\in \left[ 0,1 \right) $, finite constants $\bar{\alpha}$ and $\bar{\beta}$ such that $\mathbb{E} \left[ \mathrm{tr}(P_{k|k-1,m}) \right] \,\,\le \,\,\bar{\alpha}\bar{\rho}^k+\bar{\beta}$.
		\item[(ii)] The eavesdropper's estimation error is divergent in terms of expectation: 
		\begin{equation}\label{eq:10}
			\lim_{k\rightarrow \infty} \left\| \mathbb{E} \left[ x_k-\hat{x}_{k|k,m}^{e} \right] \right\| =\infty.
		\end{equation}
	\end{itemize}
\end{definition}

In this definition, we require the expectation of the eavesdropper's estimation error to diverge, indicating that the eavesdropper cannot accurately obtain the privacy data, specifically system states, from the intercepted information.
Meanwhile, the legitimate user obtains an unbiased estimation depending on successfully receiving and decoding the data. 
Furthermore, the legitimate user's estimation error covariance matrix remains bounded under a fading communication channel.
These conditions ensure that the PPM does not undermine the data availability for legitimate users.

\begin{problem}
	Consider a system (\ref{eq:1}) and a channel model (\ref{eq:2})-(\ref{eq4}). Design a PPM (\ref{eq:8}) and the corresponding filter (\ref{eq9}) such that the secrecy is achieved, as described in Definition \ref{definition1}.
\end{problem}
In the following sections, we will present and analyze a PPSE method, which consists of the PPM and PPF.

%\begin{remark}
%	In Definition \ref{definition1}, we introduce a concept of stochastic stability with a focus on the trace of the covariance matrix.
%	Given that $P_{k|k-1,m}\ge 0$, it follows that $\mathrm{tr} P_{k|k-1,m}\ge  \sqrt{\lambda _{\max}\left( P_{k|k-1,m}^\top P_{k|k-1,m}  \right)}$.
%	Consequently, exponential boundedness implies boundedness of $\mathbb{E} \left[ P_{k|k-1,m} \right] $, as demonstrated in studies such as \cite{Plarre_TAC2009_O} and \cite{Quevedo2013_S}.
%\end{remark}

\begin{remark}
	For the legitimate user, both the expectation and covariance are essential criteria for assessing estimation performance.
	In contrast, for the eavesdropper, divergence in the expectation is sufficient to indicate a deterioration of estimation performance, as demonstrated in \cite{Zou2023_E}.
	In complex systems, deriving a closed-form expression for the error covariance matrix can be challenging, which may hinder the analysis of data confidentiality if covariance-based definitions are employed.
	Expectation, as a key criterion for evaluating estimation performance, is generally easier to compute than covariance, due to the computational difficulties associated with higher-order moments of random variables. 
	Therefore, we adopt expectation-based confidentiality definition in order to establish an efficient and scalable framework for evaluating privacy in control systems.
\end{remark}

	\section{Privacy-Preserving State Estimation}

\subsection{Privacy-Preserving Mechanism}

As risks of privacy leakage may arise during the communication between the sensor and the remote estimator, it is inevitable to take action to protect the private data of systems. Among the existing literature about preventing privacy leakage through the wiretap channel, a class of encoding-based schemes \citep{Tsiamis2017_S, Tsiamis2018_S} is proposed by taking advantage of package loss, which works well against a powerful eavesdropper. 

An enhanced encoding scheme is developed, drawing upon the method described in \cite{Huang2022_P}. This scheme comprises two primary steps: (i) calculating the weighted difference between the innovation and its reference, and (ii) quantizing this weighted difference.
The mathematical expression of the encoding scheme is given by: 

\begin{equation}\label{eq5}
	z_k=\mathcal{Q} \left( \frac{\varepsilon _k-a^{k-t_k}\bar{\varepsilon}_{t_k}}{s} \right),
\end{equation}
where
the reference time $t_k$ is defined as 	
\begin{equation*}
	t_k=\max \left\{ t:0\le t<k, \gamma _{t,m}=1 \right\}.
\end{equation*}
It is the time slot of the last successfully received measurement at the legitimate user before $k$. 
As shown in Fig. 1, there is a reliable acknowledge channel set between the sensor and the legitimate user to make an agreement on the reference times.
The input and the output of the encoder are denoted as $\varepsilon _k\in \mathbb{R} ^{d_y}$ and $z_k\in \mathbb{R} ^{d_y}$, respectively.
The reference decoded innovation is represented as $\bar{\varepsilon}_{t_k}\in \mathbb{R} ^{d_y}$. the scaling parameters $s\ne 0$ and $a > 1$ are both scalars.

The encoding function $\mathcal{Q} \left( \cdot \right) $ operates as a stochastic uniform quantization.
To describe the characteristics of the encoding function, we define the set of the encoding levels as $\bar{\mathcal{E} }=\left\{ d\delta , d\in \mathbb{Z} , \delta >0 \right\} $. 
The parameter $\delta $ reflects the encoding quality, where a smaller value of $\delta $ corresponds to reduced data distortion.
The input of the encoding function is mapped to a specific encoding level, as described below.

For convience, we define $ \breve{\varepsilon}_k \triangleq {(\varepsilon _k -a^{k-t_k}\bar{\varepsilon}_{t_k})}/{s} \in \mathbb{R} ^{d_y}$. When $\breve{\varepsilon}_k \left( j \right) \in \left[d\delta, \left( d+1 \right) \delta \right] $, $j=1,2,...,d_y$, the encoded result follows this distribution:
\begin{equation}\label{eq6}
	\begin{cases}	\mathrm{Prob}\left\{ z_k\left( j \right) =d\delta \right\} =1-q \left(j\right),\\	\mathrm{Prob}\left\{ z_k\left( j \right) =\left( d+1 \right) \delta \right\} =q \left(j\right),\\\end{cases}	
\end{equation}
where the encoding probability is defined as	
\begin{equation*}
	q \left(j\right)=\left( \breve{\varepsilon}_k  \left( j \right)-d\delta \right) /\delta \in \left[ 0,1 \right].
\end{equation*}
In light of this, if $\breve{\varepsilon}_k \left( j \right)  $ is closer to $d\delta $ than to $\left( d+1 \right) \delta $, the value of $\breve{\varepsilon}_k \left( j \right)  $ will be equal to $d\delta $.

Let us define the quantization error as $e_k=z_k-\breve{\varepsilon}_k $, then we obtain
\begin{equation*}
	\begin{cases}	\mathrm{Prob}\left\{ e_k\left( j \right) =-q \left(j\right) \delta \right\} =1-q \left(j\right),\\	
		\mathrm{Prob}\left\{ e_k \left( j \right) =\left[ 1-q \left(j\right) \right] \delta \right\} =q \left(j\right).\\\end{cases}
\end{equation*}
The quantization error has the following statistical properties:
\begin{align*}
	&\mathbb{E} \left[ e_k  \left( j \right) \right] =0, \\
	&\mathrm{Var} \left[ e_k  \left( j \right) \right] =q \left(j\right)  \left[ 1-q \left(j\right) \right] \delta ^2 \leq \frac{\delta ^2}{4} .
\end{align*}	

The decoding mechanism corresponding to the encoding algorithm {outlined in} (\ref{eq5})-(\ref{eq6}) is derived as follows:
\begin{equation}\label{eq7}
	\bar{\varepsilon}_k=z _ks+a^{k-t_k}\bar{\varepsilon}_{t_k}.
\end{equation}
We define the decoding error as $e_{dec,k}\triangleq \bar{\varepsilon}_k-\varepsilon _k$. 
Based on the definition of the quantization error, the decoding error can be rewritten as  $e_{dec,k} = z_ks+a^{k-t_k}\bar{\varepsilon}_{t_k}-\varepsilon _k=se_k$.
Thus, the statistical properties of the decoding error are given by $\mathbb{E} \left[ e_{dec,k} \left( j \right) \right] =0$ and $\mathrm{Var}\left[ e_{dec,k} \left( j \right) \right] =s^2 \mathrm{Var} \left[ e_k  \left( j \right) \right]$,  $j=1,2,...,d_y$, which implies that it is a lossless encoding-decoding scheme in the sense of expectation.

It is assumed that the eavesdropper knows not only the system and noise parameters but also the structure and parameters of the PPM.
How can the PPM prevent a powerful eavesdropper from recovering private information?
This is achieved by exploiting the presence of missing data during transmission. 
We define a critical event as $\mathcal{E} _0=\left\{ \gamma_{k_0,m}=1,\gamma _{k_0,m}^{e}=0 \right\}$, for some $k_0\ge 0$, following the definitions provided in \cite{Tsiamis2018_S} and \cite{Huang2022_P}.
Once a critical event occurs at the time slot $k_0$, the eavesdropper will be unable to construct the decoded innovation $\bar{\varepsilon}_{k_0}$ which acts as a requisite reference for decoding any interception data $\zeta_k^e$ for $k>k_0$. 
This situation increases the eavesdropper's decoding error and deteriorates its estimation performance.
Furthermore, the introduction of the scaling parameter $a$ accelerates the degradation of the eavesdropper's estimation error, ultimately leading to the divergence of the eavesdropper's estimation.
%Further analysis of privacy-preserving performance will be provided in the next section.

\begin{remark}
	While the critical event is occurring for some $k_0\ge 0$, the eavesdropper loses data at $k_0$ and the capability to realize precise state estimation. Different from the encoding mechanism in \cite{ Huang2022_P}, the scalar $a>1$ in the proposed PPM is adopted for amplifying the effect of the data missing on the eavesdropper's estimation performance, which has a benefit that the eavesdropper's estimation error ultimately becomes unbounded in the sense of expectation. Compared with the design in \cite{Tsiamis2017_S, Tsiamis2018_S} using a matrix as the amplifying parameter, the scheme using the scalar $a$ is computationally cheaper by preventing the calculation of matrix exponentiation and inversion. 
\end{remark}

\begin{remark}
	Given an infinite time horizon, the critical event is guaranteed to occur. However, in industrial applications, it is often necessary for the critical event to happen within finite time horizon. 
	To address this, we can adopt the complementary strategy proposed in \cite{Tsiamis2020_S}.
	This complementary strategy involves employing a computationaly expensive encryption method at the initial moment, ensuring that the eavesdropper is unable to decrypt the data, thereby yielding loss of information needed to construct the reference innovation for subsequent moments.  
	Afterward, our computationally inexpensive encoding scheme takes over the privacy-preserving task in the future.
\end{remark}

%\begin{remark}
%	The PPM given in (\ref{eq5})-(\ref{eq7}) demonstrates a low computational cost and enhances transmission efficiency by reducing data volume.
%	Encoding and transmitting the innovation $\varepsilon _k$ instead of the measurement $y _k$ will effectively decrease data volume. 
%	However, during the encoding process, the calculation of the weighted difference $\varepsilon _k-a^{k-t_k}\bar{\varepsilon}_{t_k}$  necessitates a power operation on the weight $a>1$, which may inadvertently increase data length.
%	To mitigate this issue, the non-zero scalar $s$ is used to adjust data volume. Consequently, even when the value of $\varepsilon _k-a^{k-t_k}\bar{\varepsilon}_{t_k}$ becomes extremely large, the size of the data set remains manageable after encoding.
%\end{remark}

	\subsection{Privacy-Preserving Filter}

The encoding packets are transmitted through the fading channel (\ref{eq:2})-(\ref{eq4}). Only when the reception at the legitimate side is successful, will decoding be proceeded as shown in (\ref{eq7}). Therefore, the measurement noise in the dynamics of the remote filter can be mathematically modelled as $\bar{v}_{k,m} \sim \mathcal{N} \left( 0,\gamma _{k,m}R+\left( 1-\gamma _{k,m} \right) \sigma ^2I_{d_y} \right) $, similar to the model described in \cite{Sinopoli2004_K}. A pseudo-observation model with given statistics is adopted to generate the Kalman filter (KF). When the measurement fails to be received, the standard deviation is regarded as $\sigma \rightarrow \infty $ with $\gamma _{k,m}=0$.
The recursive PPSE is given in the following Theorem 1. 	

\begin{theorem}
	Consider an estimator defined as (\ref{eq9}), where the data for measurement update is generated by a dynamic system (\ref{eq:1}), encoded by a PPM (\ref{eq5})-(\ref{eq6}) and transmitted over a MFC (\ref{eq:2})-(\ref{eq4}). 
	The channel state changes from $\rho_{k-1} = \theta_n$ to $\rho_{k} = \theta_m$.
	The mode-dependent state prediction and estimation, along with their corresponding error covariance matrices, are shown as follows:
	\begin{equation}\label{eq11}			
		\begin{aligned}
			&\hat{x}_{k|k-1,n}=A\hat{x}_{k-1|k-1,n},\\
			&\hat{x}_{k|k,m}~~~=\hat{x}_{k|k-1,n}+\gamma _{k,m}K_{k,m}\bar{\varepsilon}_k,\\
			&P_{k|k-1,n}=A P_{k-1|k-1,n} A^\top+Q,\\
			&P_{k|k,m}~~~=P_{k|k-1,n}-\gamma_{k,m} K_{k,m}( CP_{k|k-1,n}C^\top+R ) K_{k,m}^\top \\
			&~~~~~~~~~~~~~~~~+\gamma _{k,m}s^2K_{k,m} R_e K_{k,m}^\top,
		\end{aligned}			
	\end{equation}
	where $R_e= \mathrm{diag} \left\{\mathrm{Var} \left[ e_k  \left( 1 \right) \right], \mathrm{Var} \left[ e_k  \left( 2 \right) \right], \dots, \mathrm{Var} \left[ e_k  \left( d_y \right) \right]   \right\}$
	and the estimator gain is 
	\begin{equation}\label{eq12}
		K_{k,m}=P_{k|k-1,n}C^\top\left( C P_{k|k-1,n} C ^\top + R \right)^{-1}.
	\end{equation}
\end{theorem}
	\begin{pf}
	Since $\sigma $-algebra generated by $\left\{ \boldsymbol{\zeta} _{1:k-1},\gamma _{k,m},z_k \right\} $ is a sub-$\sigma $-algebra generated by $\left\{ \boldsymbol{\zeta} _{1:k-1}, \gamma _{k,m},\varepsilon _k \right\} $, the following formula is obtained
	\begin{equation}\label{eq13}
		\hat{x}_{k|k,m}=\mathbb{E} \left\{ \mathbb{E} \left[ x_k | \boldsymbol{\zeta} _{1:k-1},\gamma _{k,m},\varepsilon _k \right] |\boldsymbol{\zeta} _{1:k-1},\gamma _{k,m},z_k \right\}.
	\end{equation}
	The inner conditional expectation of the above equation can be derived based on the KF as follows:
%	\begin{align}\label{eq14}
%		\hat{x}_{k|k,m}^{KF} & \triangleq \mathbb{E} [ x_k| \boldsymbol{\zeta} _{1:k-1},\gamma _{k,m},\varepsilon _k ] \nonumber \\
%		& = \mathbb{E} [ x_k| \boldsymbol{\zeta} _{1:k-2}, \zeta_{k-1,n}, \gamma _{k,m},\varepsilon _k ] \nonumber \\
%		& =\hat{x}_{k|k-1,n}+\bar{K}_{k,m} \varepsilon_k,
%	\end{align}
	\begin{align}\label{eq14}
	\hat{x}_{k|k,m}^{KF} & \triangleq \mathbb{E} [ x_k| \boldsymbol{\zeta} _{1:k-1},\gamma _{k,m},\varepsilon _k ] 
	 = \mathbb{E} [ x_k| \boldsymbol{\zeta} _{1:k-2}, \zeta_{k-1,n}, \gamma _{k,m},\varepsilon _k ] \nonumber \\
	& =\hat{x}_{k|k-1,n}+\bar{K}_{k,m} \varepsilon_k,
\end{align}
	where 
	\begin{align}\label{eq15}
		\bar{K}_{k,m}=& P_{k|k-1,n} C ^\top \nonumber \\
		& \times ( C P_{k|k-1,n} C ^\top + \gamma _{k,m} R  +( 1-\gamma _{k,m} ) \sigma ^2I_{d_y} ) ^{-1}.
	\end{align}
	By recalling that $\mathbb{E} \left[ \varepsilon _k|\boldsymbol{\zeta} _{1:k-1},\gamma _{k,m},z_k \right] =\bar{\varepsilon}_k$ and $\hat{x}_{k|k-1,m}$ is measurable in $\sigma $-algebra generated by $\left\{ \boldsymbol{\zeta} _{1:k-1},\gamma _{k,m},\varepsilon _k \right\} $, the state estimation is derived as follows:
	\begin{align}\label{eq16}
		\nonumber \hat{x}_{k|k,m}&=\mathbb{E} \left\{ \hat{x}_{k|k-1,n}+\bar{K}_{k,m}\varepsilon _k | \boldsymbol{\zeta} _{1:k-1}, \gamma _{k,m},z_k \right\} \\
		&=\hat{x}_{k|k-1,n}+\bar{K}_{k,m} \bar{\varepsilon}_k.
	\end{align}
	Similarly, the state error covariance matrix for measurement update is obtained:
	\begin{align}\label{eq17}
		& P_{k|k,m} \nonumber \\
		& =\mathbb{E} \biggl\{ \mathbb{E} \big[ \left( x_k-\hat{x}_{k|k,m} \right) \left( x_k-\hat{x}_{k|k,m} \right) ^\top \bigg| \boldsymbol{\zeta} _{1:k-1}, \gamma _{k,m}, \varepsilon _k \big] \biggr\} \nonumber \\
		& =\mathbb{E} \bigg[ \left( x_k-\hat{x}_{k|k,m}^{KF} \right) \left( x_k-\hat{x}_{k|k,m}^{KF} \right) ^\top \bigg| \boldsymbol{\zeta} _{1:k-1}, \gamma _{k,m},\varepsilon _k \bigg] + \nonumber \\
		& ~~~~ \mathbb{E} \bigg[ ( \hat{x}_{k|k,m}^{KF}-\hat{x}_{k|k,m} ) ( \hat{x}_{k|k,m}^{KF}-\hat{x}_{k|k,m} ) ^\top \bigg| \boldsymbol{\zeta} _{1:k-1}, \gamma _{k,m},\varepsilon _k \bigg] \nonumber \\
		&  =-\bar{K}_{k,m} \bigg[ CP_{k|k-1,n}C^\top+\gamma _{k,m}R+( 1-\gamma _{k,m} ) \sigma ^2I_{d_y} \bigg] \bar{K}_{k,m}^\top  \nonumber \\
		& ~~~~ +\mathbb{E} \bigg[ \bar{K}_{k,m}( \bar{\varepsilon}_k-\varepsilon _k ) ( \bar{\varepsilon}_k-\varepsilon _k ) ^\top \bar{K}_{k,m}^\top \bigg] +P_{k|k-1,n}.
	\end{align}
	When the fading occurs and the encoded data are not available, the gain matrix becomes $K_{k,m} = \lim_{\sigma \rightarrow \infty} \bar{K}_{k,m} $ with the form of (\ref{eq12}).
	Similarly, the state estimation and the error covariance matrix for updating are derived as (\ref{eq11}).
$\hfill \blacksquare$
\end{pf}

		\begin{remark}
	It should be noted that, due to the presence of PPM, the decoded innovation may not remain Gaussian. However, as indicated in \cite{Ribeiro_TSP2006_S,You_IFAC2008_M,You_ACC2009_R}, as the number of encoding levels increases, the distribution of state prediction tends to approximate Gaussian more closely. When employing a high-resolution encoding algorithm with appropriately chosen encoding levels, the decoding error covariance will tend to be zero. Simulation results provide empirical support, indicating that this approximation remains valid to a significant extent.
\end{remark}

Moreover, the computational expense of the PPF is comparable to that of the standard KF. Therefore, it is affordable for the systems or devices with some capacity constraints such as computing power and energy consumption.

\section{Performance Analysis}

\subsection{Estimation Performance for the Legitimate User}

According to the property of optimal reproduction decoders shown in \cite{Gray1998_Q}, when no data is missing, the decoded output of the PPM proposed in this paper is an unbiased estimator of the original signal input to the encoder, i.e., $\mathbb{E} \left[ \varepsilon _k-\bar{\varepsilon}_k|\gamma _{k,m}=1 \right] =0$. This indicates that the state estimation produced by the PPF is also expectational unbiased, i.e.,  $\mathbb{E} \left[ x_k-\hat{x}_{k|k,m}|\gamma _{k,m}=1 \right] =0$.

In the preceding subsection, to achieve a recursive and analytically determined form, the approximate estimation covariance $P_{k|k,m}$ is derived based on Gaussian approximation of the state prediction. However, as $\sigma \rightarrow \infty$, the actual estimation covariance is given by:
\begin{align}\label{eq18}
	& \lim_{\sigma \rightarrow \infty}   \mathbb{E} \biggl\{ \left( x_k-\hat{x}_{k|k-1,n}-\bar{K}_{k,m}\bar{\varepsilon}_k \right)  ( x_k-\hat{x}_{k|k-1,n}  \nonumber \\
	& ~~~~~~~~~~~~  -\bar{K}_{k,m}\bar{\varepsilon}_k ) ^{\top} \biggr\} \nonumber \\
	& = \Sigma_{k|k-1,n}+\gamma _{k,m}K_{k,m}\mathbb{E} \left( C \Sigma_{k|k-1,n}C^{\top}+R \right) K_{k,m}^{\top} \nonumber \\
	& ~~~~~+\gamma _{k,m} s^2 K_{k,m} R_e K_{k,m} ^{\top}  \nonumber \\ 
	& ~~~~~+\gamma _{k,m} s K_{k,m} \mathbb{E} \left[ v_k e_{k}^{\top}+e_k v_{k} ^{\top} \right] K_{k,m}^{\top}.
\end{align}	
We denote the actual estimation covariance as $\Sigma _{k|k,m}$ and then the prediction covariance is $\Sigma _{k+1|k,m}=A \Sigma _{k|k,m} A^{\top} +Q$. From the actual estimation covariance matrix, we can observe that the introduction of the PPM not only adds the decoding error covariance $s^2R_e$, but also induces correlation between the measurement noise and the quantization error, $\mathbb{E} \left[ v_ke_{k}^{\top} \right] $. Coupled with packet dropout caused by the MFC, these effects attributed to the PPM on the actual estimation covariance may pose risks to the stability of the PPF, particularly in highly unstable systems.
However, these risks can be mitigated by carefully designing the parameters of the PPM. Increasing the number of the encoding levels can reduce distortion caused by encoding, with the number of the encoding level being governed by the parameter $\delta$. Consequently, the side effects resulting from encoding-decoding errors can be alleviated by appropriately selecting the parameter $\delta$.
To better analyze the stability of the PPF, it is essential to focus on the actual covariance $\Sigma _{k|k,m}$ rather than the approximate covariance $P_{k|k,m}$. 
For notation simplicity, we henceforth denote $\Sigma _{k|k,m}$ and $\Sigma _{k+1|k,m}$ by $\Sigma _{k|k}$ and $\Sigma _{k+1|k}$, respectively.

Because of the missing data induced by the MFC, the formulations of measurement update in the PPF algorithm become functions of the random variable $\gamma_{k,m}$. 
When the system parameter $A$ is stable, the expected state estimation error covariance is always bounded. However, when $A$ is unstable, there is a risk of being divergent for the filter, especially for the case where the data is always missing or the number of encoding level is very small. Hence, we only consider the conditions for the stability of the PPF when $A$ is unstable. The following theorem gives conditions for stochastic stability of the actual mean estimation error covariance $ \mathbb{E} \left[\mathrm{tr}(\Sigma _{k|k}) \right] $ in the case where $A$ is unstable.

Before presenting the theorem, we first introduce the modified algebraic Riccati equation (MARE) defined as follows:
\begin{align}\label{eq22mod}
	\nonumber g_{\lambda}( X ) & = -\lambda AXC^\top ( CXC^\top +R ) ^{-1}CXA^\top   \\
	&~~~~+AXA^\top +Q, ~\lambda \in [ 0,1 ],
\end{align}
and it has been established that the above MARE converges in \cite{Sinopoli2004_K}.
The following theorem shows the sufficient conditions for the stability of the mean covariance matrix.

\begin{theorem}
    Consider the system (\ref{eq:1}) is unstable. The system outputs adopt the PPM (\ref{eq5})-(\ref{eq7}) and are delivered through a MFC (\ref{eq:2})-(\ref{eq4}) to a remote filter (\ref{eq11}). 
	There exists such positive scalars $\eta $, $\delta _N\in (0,1)$ and a non-negative real number $\mathcal{U} $ that the trace of the expected state estimation error covariance is bounded by
	\begin{equation}\label{eq23mod}		
		\mathbb{E} \left[\mathrm{tr}(\Sigma _{k|k}) \right] \le \alpha ^{k-1}\mathrm{tr}(\Sigma _0)+\beta \frac{1-\alpha ^{k-1}}{1-\alpha},		~ \alpha \in \left[ 0,1 \right) 
	\end{equation}
	where  $\beta \triangleq \max_{n\in \left\{ 1,2,..,M \right\}} \left\{ \sum_{j=1}^M{\mu _{k,j}p_{nj}} \right\} \mathrm{tr}\left( Q \right) +\mathcal{U} $, if the following conditions are satisfied:
	\begin{equation}\label{eq24mod}		
		\delta _N+s\eta +\frac{\delta _N}{s\eta}\in \left[ 0,1-\bar{\lambda} \right) ,			
	\end{equation}
	and
	\begin{equation}\label{eq25mod}		
		\left( \sum_{j=1}^M{\mu _{k,j}p_{nj}} \right) \left\| A \right\| ^2\in \left[ 0,1 \right) ,	
	\end{equation}
	where $\bar{\lambda}=\mathrm{arg}  \inf_{\lambda} \left\{ X|X>g_{\lambda}(X) \right\} $ and $\bar{\lambda}\in \left( 0,1 \right) $.
\end{theorem}

\begin{pf}
	For the unstable system parameter $A$, the inequality $\Sigma _{k|k} \leq A \Sigma _{k|k} A^\top +Q \triangleq \Sigma _{k+1|k}$ holds. Therefore, to establish the boundedness of the mean estimation error covariance matrix $\mathbb{E} \left[ \Sigma _{k|k} \right] $, it suffices to demonstrate that the mean prediction error covariance matrix $\mathbb{E} \left[ \Sigma _{k+1|k} \right] $is also bounded. Based on the expression of the actual estimation error covariance matrix (\ref{eq18}), we can rewrite the one-step prediction error covariance matrix as follows:
	\begin{align}\label{eq26mod}
		\nonumber	\Sigma _{k+1|k}&=A\Sigma _{k|k-1}A^\top +Q- \gamma_{k} A K_{k} \tilde{S}_k K_{k}^\top A^\top \\
		& ~~~ +\gamma_{k} A K_{k} \left( s^2 R_e + s \mathbb{E} \left[ v_k e_k^\top +e_k v_k^\top \right] \right) K_{k}^\top A^\top,
	\end{align}
	where $\tilde{S}_k\triangleq CA\Sigma _{k-1|k-1}A^\top C^\top +CQC^\top +R$.

	Before determining the upper bound of the one-step prediction error covariance matrix, we introduce the following relational expressions. For the PPM described in equations (\ref{eq5})-(\ref{eq7}) and a zero-mean random variable $\varepsilon _k$, it is derived from \cite{Gray1998_Q} and \cite{Xu2012_Q} that there exist a real number $\delta _N\in (0,1)$ named distortion rate, such that the covariance matrix of the decoding error has the following property:
	\begin{equation}\label{eq23}
		\mathbb{E} \left[ e_{dec,k}e_{dec,k}^\top \right] =s^2\mathbb{E} \left[ e_ke_{k}^\top \right] \le \delta _N\mathbb{E} \left[ \varepsilon _k\varepsilon _{k}^\top \right],
	\end{equation}
	where $\delta _N\rightarrow 0$ as the number of the encoding level  $N\rightarrow \infty $.
	
	As follows from a basic inequality $xy^\top +yx^\top \le \eta xx^\top +\eta ^{-1}yy^\top $ where $\eta $ is a positive scalar, we have  
	\begin{align}\label{eq24}
		\nonumber & \mathbb{E} \left[ v_ke_{k}^\top +e_kv_{k}^\top \right] \\
		\nonumber & \le \eta \mathbb{E} \left[ v_kv_{k}^\top \right] +\eta ^{-1}\mathbb{E} \left[ e_ke_{k}^\top \right]  =\eta R+\eta ^{-1}R_e\\
		& \le \eta \tilde{S}_k+\eta ^{-1}\left( s^{-2}\delta _N \right) \tilde{S}_k=\left( \eta +\frac{\delta _N}{s^2\eta} \right) \tilde{S}_k.
	\end{align}
	
	By substituting the inequations (\ref{eq23}) and (\ref{eq24}) into the equation (\ref{eq26mod}), we have
	\begin{align}\label{eq25}
		& \Sigma_{k+1|k} \nonumber \\
		& \le  ~A \Sigma _{k|k-1} A^\top +Q-\gamma _{k} K_{k} \tilde{S}_k K_{k}^\top A^\top \nonumber \\
		& ~~~~~ +\gamma _{k} A K_{k} \left[ \delta _N \tilde{S}_k +s \left( \eta +\frac{\delta _N}{s^2\eta} \right) \tilde{S}_k \right] K_{k}^\top A^\top \nonumber  \\
		&= ~A\Sigma _{k|k-1}A^\top +Q  \nonumber   \\
		&  ~~~~~  - \gamma _{k} \left[ 1- \left( \delta _N+s\eta +\frac{\delta _N}{s\eta} \right) \right] A K_{k}\tilde{S}_k K_{k}^\top A^\top.
	\end{align}
	
	Following the established inequality, we can determine the upper bound of the trace of mean covariance matrix based on the right-hand side of this inequality. First, we define a composite process $\left\{ G_k \right\} _{\mathbb{N} _0}$ as $G_k\triangleq \left( \Sigma _{k|k-1},\rho _{k-1} \right) $, where the channel state $\left\{ \rho _k \right\} _{\mathbb{N} _0}$ is Markovian, and the dropout process $\left\{ \gamma _k \right\} _{\mathbb{N} _0}$ is independent for given channel states. Consequently, the distribution of the matrix $\Sigma _{k|k-1}$ satisfies the  condition for all $k\in \mathbb{N} _0$:
	\begin{equation}\label{eq30mod}
		\mathrm{Prob}\left\{ \Sigma _{k|k-1}|\rho _{k-1},\rho _{k-2},... \right\} =\mathrm{Prob}\left\{ \Sigma _{k|k-1}|\rho _{k-1} \right\} .
	\end{equation}
	Therefore, the process $\left\{ G_k \right\} _{\mathbb{N} _0}$ also constitutes a Markov chain. Define the trace of covariance matrix via
	\begin{equation}\label{eq31mod}
		L_k\triangleq \mathrm{tr} (\Sigma _{k|k-1}).
	\end{equation}
	According to this definition, $L_k$ is non-negative for all $k\in \mathbb{N} _0$.
	Recall the dropout process $\left\{ \gamma _k \right\} _{\mathbb{N} _0}$, we can express the mean  covariance matrix trace based on the chain rule:
	\begin{align}\label{eq32mod}
		\mathbb{E} \left[ L_{k+1}|G_k \right] = & \mathrm{Prob}\left\{ \gamma _k=1|G_k \right\} \mathbb{E} \left[ L_{k+1}|G_k,\gamma _k=1 \right] \\
		\nonumber & +\mathrm{Prob}\left\{ \gamma _k=0|G_k \right\} \mathbb{E} \left[ L_{k+1}|G_k,\gamma _k=0 \right] .
	\end{align}
	
	We next analyze the cases of $\gamma _k=1$ and $\gamma _k=0$, respectively.
	
	(i) For $\gamma _k=1$, the error covariance matrix is bounded by 	
	\begin{align}\label{eq33mod}
		\Sigma _{k+1|k} \le & A\Sigma _{k|k-1}A^{\top} +Q \nonumber \\
		 & -\left[ 1-\left( \delta _N+s\eta +\frac{\delta _N}{s\eta} \right) \right] AK_{k}\tilde{S}_k K_{k}^{\top}A^{\top} \nonumber \\
		  \triangleq &  U_k .
	\end{align}
	Next, we need to prove that $U_k$ is bounded.
	Recall the MARE (\ref{eq22mod}). Based on [Theorem 1 to 6 in \cite{Sinopoli2004_K}], there exists a critical value $\lambda _c$ such that $U_k\le \bar{U}_k$, for $\lambda \in \left( \left. \lambda _c,1 \right] \right. $ and any initial condition $\Sigma _0\ge 0$. The upper bound of $\lambda _c$ can be determined by $\mathrm{arg} \mathop {\mathrm{inf}} \limits_{\lambda}\,\,\left\{ X|X>g_{\lambda}(X) \right\} $. Additionally, the upper bound $U_k$ can be calculated by using $\bar{U}_k=g_{\lambda}\left( \bar{U}_k \right) $ which represents the unique positive-semidefinite fixed point of the MARE. Thus, for $\gamma _k=1$ and $A$ is unstable, if
	\begin{equation}\label{eq34mod}
		\left[ 1-\left( \delta _N+s\eta +\frac{\delta _N}{s\eta} \right) \right] \in \left( \bar{\lambda},1 \right]  ,
	\end{equation}
	then the error covariance matrix is bounded as
	\begin{equation}\label{eq35mod}
		\Sigma _{k+1|k}\le \bar{U}_k\le \left( \lambda _{\max}\left\{ \bar{U}_k \right\} \right) I_{d_x }\triangleq \frac{\mathcal{U}}{d_x}I_{d_x}. 
	\end{equation}
	Hence,we derive that the first item of the equation (\ref{eq32mod}) is bounded by
	\begin{align}\label{eq36mod}
		\nonumber & \mathrm{Prob}\left\{ \gamma _k=1|G_k \right\} \mathbb{E} \left[ L_{k+1}|G_k,\gamma _k=1 \right]  \\
		\nonumber & \le  \mathrm{Prob}\left\{ \gamma _k=1|G_k \right\} \mathbb{E} \left[ \mathrm{tr}(\bar{U}_k) \right]\\
		& = \mathrm{tr}(\mathbb{E} \left[ \bar{U}_k \right] )\le \mathcal{U} .
	\end{align}	
	(ii) For $\gamma _k=0$, we first derive the bound on the trace of the covariance matrix:
%	\begin{align}\label{eq37mod}	
%		\nonumber & \mathbb{E} \left[ L_{k+1}|G_k=\left( \Sigma ,\theta _n \right) ,\gamma _k=0 \right] \\
%		\nonumber & =\mathbb{E} \left[ \mathrm{tr}\left( A\Sigma A^{\top} +Q \right) \right] \\
%		\nonumber & =\mathrm{tr}\left( A\Sigma A^{\top}  \right) +\mathrm{tr}\left( Q \right) \\
%		& \le \left\| A \right\| ^2\mathrm{tr}{(\Sigma)} +\mathrm{tr}\left( Q \right).
%	\end{align}	
	\begin{align}\label{eq37mod}	
			& \mathbb{E} \left[ L_{k+1} | G_k=\left( \Sigma ,\theta _n \right) ,\gamma _k=0 \right] 
			  = \mathbb{E} \left[ \mathrm{tr} \left( A \Sigma A ^{\top} + Q \right) \right] \nonumber \\
			& =\mathrm{tr} \left( A\Sigma A ^{\top}  \right) + \mathrm{tr} \left( Q \right) 
			    \le \left\| A \right\| ^2\mathrm{tr} {(\Sigma)} +\mathrm{tr} \left( Q \right).
	\end{align}	
	Next, considering the independence of the dropout process $\left\{ \gamma _k \right\} _{\mathbb{N} _0}$ and the Markovian properties of the channel state $\left\{ \rho _k \right\} _{\mathbb{N} _0}$, we calculate the probability as follows:
	\begin{align}\label{eq38mod}	
		\nonumber & \mathrm{Prob} \left\{ \gamma _k =0|G_k \right\}\\
		\nonumber & =\sum_{j=1}^M{\mathrm{Prob}\left\{ \gamma _k=0|\Sigma _{k|k-1}=\Sigma , \rho _{k-1}=\theta _n,\rho _k=\theta _j \right\}}\\
		\nonumber &  ~~~~~~~~ \times \mathrm{Prob}\left\{ \rho _k=\theta _j|\Sigma _{k|k-1}=\Sigma , \rho _{k-1}=\theta _n \right\} \\
		& =\sum_{j=1}^M{\mu _{k,j}p_{nj}}.
	\end{align}

	Substituting equations (\ref{eq36mod}) - (\ref{eq38mod}) into (\ref{eq31mod}), we obtain
	\begin{align}\label{eq39mod}	
		& \mathbb{E} \left[ L_{k+1}|G_k=\left( \Sigma ,\theta _n \right) \right] \\
		\nonumber & \le \mathcal{U} +\left( \sum_{j=1}^M{\mu _{k,j}p_{nj}} \right) \left[ \left\| A \right\| ^2\mathrm{tr}(\Sigma) +\mathrm{tr}\left( Q \right) \right] \\
		\nonumber & \le \left( \sum_{j=1}^M{\mu _{k,j}p_{nj}} \right) \left\| A \right\| ^2L_k+\left( \sum_{j=1}^M{\mu _{k,j}p_{nj}} \right) \mathrm{tr}\left( Q \right) +\mathcal{U}.		
	\end{align}
	%			where $\delta _N+s\eta +\frac{\delta _N}{s\eta}\in \left[ 0,1-\bar{\lambda} \right) $.	
	Utilizing the stochastic Lyapunov function approach from \cite{Quevedo2013_S} gives us:
	\begin{align}\label{eq40mod}	
		\nonumber & \mathbb{E} \left[ L_{k+1} \mid G_k = \left( \Sigma ,\theta _n \right) \right] \\
		\nonumber & \le \left( \sum_{j=1}^M {\mu _{k,j}p_{nj}} \right) \left\| A \right\| ^2 L_k
		+ \left( \sum_{j=1}^M {\mu _{k,j}p_{nj}} \right) \mathrm{tr}\left( Q \right) + \mathcal{U} \\
		& \le \alpha L_k + \beta,
	\end{align}
	where 	
	\begin{equation}\label{eq41mod}	
		\begin{aligned}
			& \alpha \in \left[ 0,1 \right), \\
			& \beta \triangleq \max_{n\in \left\{ 1,2,\ldots,M \right\}} \left\{ \sum_{j=1}^M {\mu_{k,j}p_{nj}}  \right\}  \, \mathrm{tr}(Q) + \mathcal{U}\ge 0.		
		\end{aligned}
	\end{equation}
	The equation (\ref{eq40mod}) is valid for all $G_k=\left( \Sigma _{k|k-1},\rho _{k-1} \right) $. 
	Considering the Markovian properties of the process $\left\{ G_k \right\} _{k\in \mathbb{N} _0}$, we can obtain the following inequality, based on Proposition 3.2 in \cite{Meyn_SIAM_JCO1989_E}:
%	\begin{align}\label{eq42mod}
%		\nonumber \mathbb{E} \left[ L_k \mid G_0 = G \right] & \le \alpha^k L_0 + \beta \sum_{t=0}^{k-1} \alpha^t \\
%		& \le \alpha^k L_0 + \beta \frac{1 - \alpha^k}{1 - \alpha}.
%	\end{align}	
	\begin{align}\label{eq42mod}
	\mathbb{E} \left[ L_k \mid G_0 = G \right] & \le \alpha^k L_0 + \beta \sum_{t=0}^{k-1} \alpha^t 
	\le \alpha^k L_0 + \beta \frac{1 - \alpha^k}{1 - \alpha}.
    \end{align}			 
	Thus, given the initial condition $L_0$, it follows that (\ref{eq23mod}) is satisfied under the conditions for the encoder and channel (\ref{eq24mod})-(\ref{eq25mod}). 
$\hfill \blacksquare$
\end{pf}

Theorem 2 demonstrates the boundedness conditions of the actual state error covariance for the PPF, which provides guidelines on the design of the number of encoding level and the scaling parameter. It is noticeable that the plants may meet certain conditions, for example, the outputs are scalar measurements or each element of the output vector has no correlation. In these cases, the boundedness conditions can be relaxed, which is shown in the following corollary. 

\begin{corollary}
	Consider an unstable system (\ref{eq:1}) with the outputs satisfying $\mathrm{Cov} \left( y_k\left( i \right) ,y_k\left( j \right) \right) =0,\forall i\ne j\in \left\{ 1,2,...,d_y \right\} $. The PPM (\ref{eq5})-(\ref{eq7}) is carried out before the outputs are transmitted to a remote filter (\ref{eq11}) over a MFC (\ref{eq:2})-(\ref{eq4}). With a real number $\delta _N\in \left( 0,1 \right) $ and a non-negative real number $\mathcal{U} $, the trace of the expected state estimation error covariance is bounded by (\ref{eq23mod}), if
	\begin{equation}\label{eq30-1}						
		\delta _N+ 2 \sqrt{\delta _N}\in \left[ 0,1-\bar{\lambda} \right) ,		
	\end{equation}
	and
	\begin{equation}\label{eq30-2}						
		\left( \sum_{j=1}^M{\mu _{k,j}p_{nj}} \right) \left\| A \right\| ^2\in \left[ 0,1 \right) ,				
	\end{equation}
	where $\bar{\lambda}=\mathrm{arg} \mathop {\mathrm{inf}} \limits_{\lambda} \left\{ X|X>g_{\lambda}(X) \right\} $ and $\bar{\lambda}\in \left( 0,1 \right) $. 
\end{corollary}
\begin{pf}
	The proof is analogous to that of Theorem 2, with the primary distinction being the covariance of the measurement noise and the encoding error, since the outputs are scalar and each element of the output vector is uncorrelated. Using the conclusions from the proof of Theorem 2, along with the Cauchy-Schwarz inequality, we can derive the following inequality: 
	\begin{align}\label{eq31}
		\nonumber &	\mathbb{E} \left[ v_k\left( j \right) e_k\left( j \right)  \right] \\
		\nonumber & \le \sqrt{\mathbb{E} \left[ v_k\left( j \right) v_k\left( j \right) \right]}\cdot \sqrt{\mathbb{E} \left[ e_k\left( j \right) e_k\left( j \right) \right]}\\
		\nonumber	& \le \sqrt{R\left( j,j \right)}\cdot \sqrt{s^{-2}\delta _N\tilde{S}_k\left( j,j \right)}\\
		& \le \frac{\sqrt{\delta _N}}{s}\tilde{S}_k\left( j,j \right).
	\end{align}
	Therefore,  for $\gamma _k=1$, the error covariance matrix is bounded by
	\begin{align}\label{eq32}
		\nonumber \Sigma _{k+1|k} \le & A\Sigma _{k|k-1}A^{\top} +Q\\
		& -\left[ 1-\left( \delta _N+ 2 \sqrt{\delta _N} \right) \right] A K_{k}\tilde{S}_kK_{k}^{\top}A^{\top} \triangleq V_k .
	\end{align}
	The condition that $V_k$ to be bounded is 
	\begin{equation}\label{eq46mod}
		\left[ 1-\left( \delta _N+2 \sqrt{\delta _N} \right) \right] \in \left( \bar{\lambda},1 \right] . 
	\end{equation}
	The steps for calculating the upper bound of the error covariance matrix remain consistent with those outlined in the proof of Theorem 2. The remainder of the proof of Corollary 1 follows the same logic as that of Theorem 2. 
	$\hfill \blacksquare$
\end{pf}
Since packet dropout arises from channel conditions and the measurement forms primarily influence the encoding results, variations in the measurement forms chiefly affect the boundedness condition regarding the encoding aspect.

\subsection{Privacy-Preserving Performance}

In this subsection, we are going to analyze the performance of the PPM (\ref{eq5})-(\ref{eq6}) by considering the state estimation performance at the eavesdropper's side. For the eavesdropper with the wiretap-channel states $ \rho_{k}^e = \theta_n$ and $ \rho_{k+1}^e = \theta_m$, let us define the one-step prediction error and state estimation error as $\tilde{x}_{k+1|k,n}^{e}=x_{k+1}-\hat{x}_{k+1|k,n}^{e}$ and $\tilde{x}_{k+1|k+1,m}^{e}=x_{k+1}-\hat{x}_{k+1|k+1,m}^{e}$. Then, the dynamics of the one-step prediction error and the state estimation error is shown as follows
\begin{equation}\label{eq33}
	\left\{
	\begin{aligned}
		\tilde{x}_{k+1|k,n}^{e}&=A\tilde{x}_{k|k,n}^{e},\\ 
		\tilde{x}_{k+1|k+1,m}^{e}&=\tilde{x}_{k+1|k,n}^{e}+\gamma _{k+1,m}^{e}K_{k+1,m}^{e}\bar{\varepsilon}_{k+1}^{e} \\
		&=\tilde{x}_{k+1|k,n}^{e}+\gamma _{k+1,m}^{e}K_{k+1,m}^{e}\left( \varepsilon _{k+1}+\bar{e}_{k+1} \right),
	\end{aligned}
	\right.
\end{equation}
where $\bar{e}_k\triangleq \bar{\varepsilon}_{k}^{e}-\varepsilon_k$ is the decoding error at the eavesdropper's side.

By taking the expectation, we have
\begin{align}\label{eq34}
	\nonumber &\mathbb{E} \left[ \tilde{x}_{k+1|k,n}^{e} \right] =A\mathbb{E} \left[ \tilde{x}_{k|k,n}^{e} \right],\\	
	\nonumber &\mathbb{E} \left[ \tilde{x}_{k+1|k+1,m}^{e} \right] =  \bar{\gamma}_{k+1,m}^{e}K_{k+1,m}^{e}\left( \mathbb{E} \left[ \varepsilon _{k+1} \right] +\mathbb{E} \left[ \bar{e}_{k+1} \right] \right) \\
	& ~~~~~~~~~~~\ \ ~~~~~~~+\mathbb{E} \left[ \tilde{x}_{k+1|k,n}^{e} \right].
\end{align}

The mean estimation error is a crucial index for state estimation performance. We want to show that the mean error $\mathbb{E} \left[ \bar{e}_{k+1} \right] $ induced by the PPM (\ref{eq5})-(\ref{eq6}) will worsen the eavesdropper's estimation performance once a critical event occurs. At that moment, the legitimate user successfully decodes the innovation which is also utilized for updating the reference innovation at the next moment. However, with the unsuccessful interception, the eavesdropper neither achieves lossless decoding nor knows the reference information for decoding at the coming moment. Therefore, a bias is introduced in the decoded innovation used for the eavesdropper's state estimation, which further leads to biased estimation. 
To analyze how the PPM-induced error $\bar{e}_k$ worsens the eavesdropper's estimation performance caused by the critical event, firstly we consider an extreme case where all the interceptions are successful after a critical event occurs.

\begin{proposition}
	Suppose that, for some $k_0\ge 0$, the channel states are $ \rho_{k_0} = \theta_ {\bar{m}_0}$ and $ \rho_{k_0}^e =\theta_{\breve{m}_0}$, $\bar{m}_0, \breve{m}_0 \in \left\{ 1,2,\dots,M \right\}$
	and that the following two events occur simultaneously:
	\begin{align}\label{eq35}
		\mathcal{E} _0 & = \left\{ \gamma_{k_0,\bar{m}_0}=1, \gamma_{k_0, \breve{m}_0}^{e}=0 \right\} , \nonumber\\
		\mathcal{E}_1 & = \left\{ \gamma_{k,m}^{e}=1, \forall k>k_0 \right\}.
	\end{align}
	Then, under the joint event $\mathcal{E}_0 \cap \mathcal{E}_1$ and if $a > 1$ , the sequence of the PPM-induced error shown as (\ref{eq33}) is divergent, that is  $\lim_{k\rightarrow \infty} \left\| \mathbb{E} \left[ \bar{e}_k \right] \right\| =\infty $.
\end{proposition}

\begin{pf}
	When a critical event $\mathcal{E} _0$ occurs for some $k_0\ge 0$, the eavesdropper cannot recover the decoded innovation without obtaining the data transmitted from the system to the legitimate user at that moment. Then, for the next moment $k_0+1$, according to the formulation of decoding (\ref{eq7}), the eavesdropper still cannot exactly reconstruct the decoded innovation $\bar{\varepsilon}_{k_0+1}$ because of the lack of the reference decoded innovation $\bar{\varepsilon}_{k_0}$. As a result, the decoded innovation of the eavesdropper becomes biased, which means $\mathbb{E} \left[ \bar{e}_{k_0+1} \right] =\delta ^e$ for  $\delta ^e\ne 0$. 
	Based on the PPM (\ref{eq5})-(\ref{eq7}), the decoding error of the eavesdropper will propagate as follows for $k\ge k_0+1$:
	\begin{align}\label{eq36}
		\nonumber \bar{e}_k & =\bar{\varepsilon}_{k}^{e}-\varepsilon _k=\bar{\varepsilon}_{k}^{e}-\bar{\varepsilon}_k+e_ks\\
		\nonumber & =a\left( \bar{\varepsilon}_{k-1}^{e}-\bar{\varepsilon}_{k-1} \right) +e_ks\\
		\nonumber &=a\left[ a\left( \bar{\varepsilon}_{k-2}^{e}-\bar{\varepsilon}_{k-2} \right) \right] +e_ks\\
		\nonumber &\ldots\\
		\nonumber &=a^{k-k_0-1}\left( \bar{\varepsilon}_{k_0+1}^{e}-\bar{\varepsilon}_{k_0+1} \right) +e_ks\\ 
		& =a^{k-k_0-1}\left( \bar{e}_{k_0+1}-e_{k_0+1}s \right) +e_ks.
	\end{align}
	Under such propagation mode, if $a > 1$, then we have
	$\lim_{k\rightarrow \infty} \left\| \mathbb{E} \left[ \bar{e}_k \right] \right\|  =\infty$. 
	$\hfill \blacksquare$
\end{pf}

The following proposition is proposed as evidence for the eavesdropper's estimator, suggesting that in the worst-case scenario, the expectation of the estimation error for the eavesdropper becomes divergent.

\begin{proposition}
	For a system (\ref{eq:1}), with consideration of the filtering algorithm (\ref{eq11}), there exists a positive scalar $\bar{\kappa}$ such that $\left( K_{k,m}^{e} \right) ^\top K_{k,m}^{e}\ge \bar{\kappa}I_{d_y}$ holds.
\end{proposition}
\begin{pf}
	According to the gain formula of the estimator shown in (\ref{eq12}), we have
	\begin{align}\label{eq37}
		\nonumber \left( K_{k,m}^{e} \right) ^\top K_{k,m}^{e}= & \left( C P_{k|k-1,n}^{e} C^\top +R \right) ^{-1} C  P_{k|k-1,n}^{e}  \\
		& \times P_{k|k-1,n}^{e} C^\top \left( CP_{k|k-1,m}^{e}C^\top +R \right) ^{-1}	.
	\end{align}
	
	Since $P_{k|k-1,n}^{e}\ge Q\ge \lambda _{\min}\left\{ Q \right\} I_{d_x}$ and the boundedness of the innovation covariance, then the above equation becomes
	\begin{align}
		\nonumber \left( K_{k,m}^{e} \right) ^\top K_{k,m}^{e} \ge & \left( \lambda _{\min}\left\{ Q \right\} \right) ^2\lambda _{\min}\left\{ CC \right\} \\
		& \times \left( \lambda _{\max}\left\{ C P_{k|k-1,n}^{e}C^\top +R \right\} \right) ^{-2}	.
	\end{align}
	Therefore, for a positive scalar $\bar{\kappa}=(\lambda_{\min}\{ Q \} )^2\lambda_{\min}\{CC\}$ $ (\lambda_{\max}\{ C P_{k|k-1,n}^{e}C^\top +R \} ) ^{-2}$, there exists $( K_{k,m}^{e} )^\top K_{k,m}^{e}$ $ \ge \bar{\kappa}I_{d_y}$.
	$\hfill \blacksquare$
\end{pf}

With these propositions, we are now able to analyze the eavesdropper's worst-case estimation performance.  Based on the dynamics of the mean estimation error (\ref{eq34}), the eavesdropper's estimator is regarded as a bad one if the mean estimation error $\{ \mathbb{E} [ \tilde{x}_{k|k,m}^{e} ] \} _{k>0}$ is divergent, which is presented in the following lemma.

\begin{lemma}
	Consider the system in (\ref{eq:1}) with the filtering algorithm (\ref{eq11}). Suppose that for some $k_0\ge 0$, the states of authorized channel and wiretap channel are given by $ \rho_{k_0} = \theta_ {\bar{m}_0}$ and $ \rho_{k_0}^e =\theta_{\breve{m}_0}$, respectively.
	Define the events $\mathcal{E} _0=\left\{ \gamma _{k_0,\bar{m}_0}=1, \gamma _{k_0,\breve{m}_0}^{e}=0 \right\} $ and $\mathcal{E}_1 = \left\{ \gamma_{k,m}^{e}=1, \forall k > k_0 \right\}$. 
	The event $\mathcal{E}_1$ indicates that the eavesdropper successfully intecepts information for any channel states $\rho_k^e = \theta_m $ after $k_0$.
	The joint event $\mathcal{E} _0\cap \mathcal{E} _1$ occurs during transmission over the fading channel (\ref{eq:2})-(\ref{eq4}). Under the PPM in (\ref{eq5})-(\ref{eq7}), if $a > 1$, the mean estimation error $\left\{ \mathbb{E} \left[ \tilde{x}_{k|k,m}^{e} \right] \right\}_{k>0}$is divergent.
\end{lemma}

\begin{pf}
	Similar to the idea of \cite{Zou2023_E}, contradiction is utilized for the proof of this lemma. Let's assume that when $\mathcal{E} _0\cap \mathcal{E} _1$, for any $k\ge k_0$, the sequence $\left\{ \mathbb{E} \left[ \tilde{x}_{k|k,m}^{e} \right] \right\} _{k>0}$ is eventually bounded. In other words, we have $\lim_{k\rightarrow \infty} \left\| \mathbb{E} \left[ \tilde{x}_{k|k,m}^{e} \right] \right\| \le \tilde{\chi}$, where $\tilde{\chi}$ is a positive scalar.
	According to the dynamics (\ref{eq34}), for the channel state varies from $\rho_k^e = \theta_m$ to $\rho_{k+1}^e = \theta_{\breve{m}}$, we have
%	\begin{align}\label{eq39}
%		& \left\| K_{k,m}^{e}\left( \mathbb{E} \left[ \varepsilon _{k+1} \right] +\mathbb{E} \left[ \bar{e}_{k+1} \right] \right) \right\| \nonumber \\
%		& \le \left\| \mathbb{E} [ \tilde{x}_{k+1|k+1,\breve{m}}^{e} ] \right\| +\left\| A\mathbb{E} [ \tilde{x}_{k|k,m}^{e} ] \right\| \nonumber \\
%		&  \le \left( 1+\left\| A \right\| \right) \tilde{\chi}.
%	\end{align}
	\begin{align}\label{eq39}
	& \left\| K_{k,m}^{e}\left( \mathbb{E} \left[ \varepsilon _{k+1} \right] +\mathbb{E} \left[ \bar{e}_{k+1} \right] \right) \right\| 
	\le \left\| \mathbb{E} [ \tilde{x}_{k+1|k+1,\breve{m}}^{e} ] \right\| +\left\| A\mathbb{E} [ \tilde{x}_{k|k,m}^{e} ] \right\| \nonumber \\
	&  \le \left( 1+\left\| A \right\| \right) \tilde{\chi}.
   \end{align}
	From (\ref{eq37}), the following inequality is derived
	\begin{equation}\label{eq40}
		\bar{\kappa}^{\frac{1}{2}}\left\| \left( \mathbb{E} \left[ \varepsilon _{k+1} \right] +\mathbb{E} \left[ \bar{e}_{k+1} \right] \right) \right\| \le \left( 1+\left\| A \right\| \right) \tilde{\chi}.
	\end{equation}
	Consequently, $\lim_{k\rightarrow \infty} \bar{\kappa}^{\frac{1}{2}}\left\| \left( \mathbb{E} \left[ \varepsilon _{k+1} \right] +\mathbb{E} \left[ \bar{e}_{k+1} \right] \right) \right\| \le \left( 1+\left\| A \right\| \right) $ $ \tilde{\chi}$. However, the condition that $\left\| \mathbb{E} \left[ \varepsilon _{k+1} \right] \right\| =0$ and $\lim_{k\rightarrow \infty} \left\| \mathbb{E} \left[ \bar{e}_{k+1} \right] \right\| $ $  =\infty $ is contradictory to the above conclusion. Therefore, as $\lim_{k\rightarrow \infty} \left\| \mathbb{E} \left[ \tilde{x}_{k|k,m}^{e} \right] \right\| =\infty $, we can draw a conclusion that the sequence $\left\{ \mathbb{E} \left[ \tilde{x}_{k|k,m}^{e} \right] \right\} _{k>0}$ is divergent for any $k\ge k_0$ and $\mathcal{E} _0\cap \mathcal{E} _1$. 
	$\hfill \blacksquare$
\end{pf}

We accomplish the performance analysis of the state estimator at the eavesdropper's side whose results imply that the privacy is preserved for the worst case. Next, a general conclusion is drawn that the eavesdropper's estimation performance gets worse if the critical event is triggered, before which the following lemma is requisite.

\begin{lemma}
	For a probability space $\left( \Omega ,\mathcal{F} , \mathcal{P} \right) $, there exists $\mathcal{J} _1\subseteq \mathcal{F} $ being a sigma algebra $\sigma \left( \mathcal{F} \right) $ and a random variable $Y\in \,\,\Omega $ with $\mathbb{E} \left[ Y \right] \le \infty $. Then $\mathbb{E} \left[ Y|\mathcal{J} _1 \right] $ is a $\mathcal{J}_1$-measurable function satisfying $\int_{J_1} {\mathbb{E} \left[ Y| \mathcal{J} _1 \right] d \mathcal{P} }=\int_{J_1}{Y\,\,d \mathcal{P} }, J_1\in \mathcal{J} _1$. Let us define that $\mathbb{E} \left[ \tilde{Y}_{\mathcal{J} _i} \right] =\mathbb{E} \left\{ Y-\mathbb{E} \left[ Y|\mathcal{J} _i \right] \right\} , i=1,2$. Then if $\mathcal{J} _2\subseteq \mathcal{J} _1\subseteq \mathcal{F} $, we have 
	\begin{equation}\label{eq41}
		\left\| \mathbb{E} \left[ \tilde{Y}_{\mathcal{J} _1}|\mathcal{J} _1 \right] \right\| \le \left\| \mathbb{E} \left[ \tilde{Y}_{\mathcal{J} _2}|\mathcal{J} _1 \right] \right\|.
	\end{equation}
\end{lemma}
\begin{pf}
	Since $\mathcal{J} _2\subseteq \mathcal{J} _1\subseteq \mathcal{F} $ and $Y-\mathbb{E} \left[ Y|\mathcal{J} _2 \right] $ is measurable with respect to $\sigma \left( \mathcal{J} _1 \right) $, by the tower property, it is derived that  
	\begin{align}\label{eq42}
		\nonumber & \bigg\| \mathbb{E} \left\{ Y-\mathbb{E} \left[ Y|\mathcal{J} _1 \right] |\mathcal{J} _1 \right\} \bigg\| -\bigg\| \mathbb{E} \left\{ Y-\mathbb{E} \left[ Y|\mathcal{J} _2 \right] |\mathcal{J} _1 \right\} \bigg\| \\
		& =0-\bigg\| \mathbb{E} \left[ Y|\mathcal{J} _1 \right] -\mathbb{E} \left[ Y|\mathcal{J} _2 \right] \bigg\|.
	\end{align}
	As $\bigg\| \mathbb{E} \left[ Y|\mathcal{J} _1 \right] -\mathbb{E} \left[ Y|\mathcal{J} _2 \right] \bigg\| \ge 0$, we have
	\begin{align}
		\bigg\| \mathbb{E} \left\{ Y-\mathbb{E} \left[ Y|\mathcal{J} _1 \right] |\mathcal{J} _1 \right\} \bigg\| -\bigg\| \mathbb{E} \left\{ Y-\mathbb{E} \left[ Y|\mathcal{J} _2 \right] |\mathcal{J} _1 \right\} \bigg\| \le 0,
	\end{align}
	that is $\bigg\| \mathbb{E} \left\{ Y-\mathbb{E} \left[ Y|\mathcal{J} _1 \right] |\mathcal{J} _1 \right\} \bigg\| \le \bigg\| \mathbb{E} \left\{ Y-\mathbb{E} \left[ Y|\mathcal{J} _2 \right] |\mathcal{J} _1 \right\} \bigg\| $. 
	$\hfill \blacksquare$
\end{pf}

Now, let us denote the channel outcomes by a batch of random variables such as $\boldsymbol{h}_{0:k}=\left( \boldsymbol{\gamma}_{0:k}, \boldsymbol{\gamma}_{0:k}^{e} \right) =\left( \gamma _{0,\bar{n}},..,\gamma _{k,\bar{m}},\gamma _{0,\breve{n}}^{e},..,\gamma _{k,\breve{m}}^{e} \right) $ 
whose values are taken from the set $\left\{ 0,1 \right\} ^{2k+2}$. 
Also denote $\boldsymbol{l}_{0:k}=\left( l_{0,\bar{n}},..,l_{k,\bar{m}},l_{0,\breve{n}}^{e},..,l_{k,\breve{m}}^{e} \right) $ is composed by any fixed element of $\left\{ 0,1 \right\} ^{2k+2}$. 
Then, based on the event that the channel outcomes $\boldsymbol{h}_{0:k}$ are exactly the values of $\boldsymbol{l}_{0:k}$, 
the information $\mathcal{I} _{k,m}^{e}$ intercepted by the eavesdropper 
is alternatively expressed by $\mathcal{J} _k \left( \boldsymbol{l}_{0:k} \right) =\left\{ \zeta _t \left( l_{t,m} \right) : l_{t,m}=1, t\le k, m\in \left\{ 1,2,...,M \right\} \right\} $. With the aforementioned descriptions, we now can shed light on the confidentiality achieved by using the proposed PPM.

\begin{theorem}
	Consider the system (\ref{eq:1}) communicating with the remote estimator (\ref{eq11}) over the fading channel (\ref{eq:2})-(\ref{eq4}) in the presence of an eavesdropper. Suppose that for some $k\ge 0$, 
	and any channel states $\theta_{\bar{m}}, \theta_{\breve{m}} \in \Theta$, the following holds:
	\begin{equation}\label{eq44}
		\mathrm{Prob}\left\{ \gamma _{k,\bar{m}}=1, \gamma _{k,\breve{m}}^{e}=0 \right\} =1.
	\end{equation}
	Then, under the protection of the PPM (\ref{eq5})-(\ref{eq7}), the mean estimation error $\{ \mathbb{E} [ \tilde{x}_{k|k,\breve{m}}^{e} ] \} _{k>0}$ is divergent, that is $\lim_{k\rightarrow \infty}$ $ \| \mathbb{E} [ \tilde{x}_{k|k,\breve{m}}^{e} ] \| =\infty $, at the eavesdropper's side.
\end{theorem}
\begin{pf}
	Let $\rho_k = \theta_{\bar{m}}$ and $\rho_k^e = \theta_{\breve{m}}$ denote the states of authorized and wiretap channels at time $k$.
	To formally describe the cases of the eavesdropper's interception results after the critical event, new channel outcome sequences are defined as follows:
	
	1) Worst case: 
	
	for $0\le k<k_0$, $\left( \breve{\gamma}_{k,\bar{m}}, \breve{\gamma}_{k,\breve{m}}^{e} \right) =\left( \gamma _{k,\bar{m}}, \gamma _{k,\breve{m}}^{e} \right) $;
	
	for $k=k_0$, $\left( \breve{\gamma}_{k,\bar{m}}, \breve{\gamma}_{k,\breve{m}}^{e} \right) =\left( 1, 0 \right) $; 
	
	for $k>k_0$, $\left( \breve{\gamma}_{k,\bar{m}}, \breve{\gamma}_{k,\breve{m}}^{e} \right) =\left( \gamma _{k,\bar{m}}, 1 \right) $.
	
	2) Other cases after the critical event: 
	
	for $0\le k<k_0$, $\left( \tilde{\gamma}_{k,\bar{m}}, \tilde{\gamma}_{k,\breve{m}}^{e} \right) =\left( \gamma _{k,\bar{m}}, \gamma _{k,\breve{m}}^{e} \right) $; 
	
	for $k=k_0$, $\left( \tilde{\gamma}_{k,\bar{m}}, \tilde{\gamma}_{k,\breve{m}}^{e} \right) =\left( 1, 0 \right) $; 
	
	for $k>k_0$, $\left( \tilde{\gamma}_{k,\bar{m}}, \tilde{\gamma}_{k,\breve{m}}^{e} \right) =\left( \gamma_{k,\bar{m}}, \gamma_{k,\breve{m}}^{e} \right) $, where $\exists \gamma _{t,\breve{m}}^{e}\ne 1$, for some $t>k_0$.
	
	Again, we define a new batch of the channel outcomes by $\boldsymbol{\breve{h}}_{0:k}=( \boldsymbol{\breve{\gamma}}_{0:k}, \boldsymbol{\breve{\gamma}}_{0:k}^{e} ) $ and $\boldsymbol{\tilde{h}}_{0:k}=\left( \boldsymbol{\tilde{\gamma}}_{0:k}, \boldsymbol{\tilde{\gamma}}_{0:k}^{e} \right) $. 
	
%	Then, the information $\mathcal{I} _{k,m}^{e}$ intercepted by the eavesdropper based on the two cases are $\mathcal{J} _k\left( \boldsymbol{\breve{h}}_{0:k} \right) =\left\{ \zeta _t \left( \breve{h}_{t,m} \right) : \breve{h}_{t,m}=1, t\le k \right\} $ 
%	and $\mathcal{J} _k\left( \boldsymbol{\tilde{h}}_{0:k} \right) = \left\{ \zeta _t \left( \tilde{h}_{t,m} \right) : \tilde{h}_{t,m}=1, t\le k \right\} $. 
	Then, the information $\mathcal{I} _{k,m}^{e}$ intercepted by the eavesdropper based on the two cases are $\mathcal{J} _k\left( \boldsymbol{\breve{h}}_{0:k} \right) = \biggl\{ \zeta _t \left( \breve{h}_{t,m} \right) : \breve{h}_{t,m}=1,$  $  t\le k \biggr\} $ 
	and $\mathcal{J} _k\left( \boldsymbol{\tilde{h}}_{0:k} \right) = $ 
	$\left\{ \zeta _t \left( \tilde{h}_{t,m} \right) : \tilde{h}_{t,m}=1, t\le k \right\} $. 
	
	According to the [Lemma 3, \cite{Tsiamis2020_S}], the sigma algebra $\sigma \left( \mathcal{J} _k\left( \boldsymbol{\breve{h}} \right) \right) $ can be described in term of $\sigma ( \mathcal{J} _k( \boldsymbol{\tilde{h}} ) ) $. 
	For $0\le k\le k_0$, the channel outcomes $ \boldsymbol{\breve{h}}_{0:k}$ are identical to $ \boldsymbol{\tilde{h}}_{0:k}$. 
	For $k>k_0$, every set $\breve{D}$ in the sigma algebra of  $\mathcal{J} _k\left( \boldsymbol{\breve{h}}_{k_0:k} \right) $, 
	that is $\breve{D}=\left\{ \mathcal{J} _k\left( \boldsymbol{\breve{h}}_{k_0:k} \right) \in \breve{F} \right\} $, 
	$\breve{F}\in \sigma \left( \mathcal{J} _k\left( \boldsymbol{\breve{h}}_{k_0:k} \right) \right) $, 
	has the form of $\breve{D}=\bigcup_{l\in \mathcal{L}}{\tilde{D}_l}$ , 
	where $\tilde{D}_l=\left\{ \mathcal{J} _k\left( \boldsymbol{\tilde{h}}_{k_0:k}=l_{k_0:k,m} \right) \in \tilde{F}_l \right\} , 
	\tilde{F}_l\in \sigma \left( \mathcal{J} _k\left( \boldsymbol{l}_{k_0:k} \right) \right) $.
	Therefore, it is obvious that $\mathcal{J} _k\left( \boldsymbol{\tilde{h}}_{0:k} \right) \subseteq \mathcal{J} _k\left( \boldsymbol{\breve{h}}_{0:k} \right) $ and $\mathcal{J} _k\left( \boldsymbol{\tilde{h}} \right) $ is $\mathcal{J} _k\left( \boldsymbol{\breve{h}} \right) $ -measurable.
	By adopting the Lemma 2, we have $ \left\| \mathbb{E} \left[ \tilde{x}_{k|k}^{e}|\mathcal{J} _k\left( \tilde{h} \right) \right] \right\| \le \left\| \mathbb{E} \left[ \tilde{x}_{k|k}^{e}|\mathcal{J} _k\left( \breve{h} \right) \right] \right\| $.
	Since the result of Lemma 1 shows that $ \lim_{k\rightarrow \infty} \| \mathbb{E} [ \tilde{x}_{k|k,\breve{m}}^{e}|\mathcal{J}_k\left( \boldsymbol{\tilde{h}} \right) ] \| =\infty $,
	there is no doubt about $\lim_{k\rightarrow \infty} \| \mathbb{E} [ \tilde{x}_{k|k,\breve{m}}^{e}|\mathcal{J} _k( \boldsymbol{\breve{h}} ) ] \| =\infty $.
	Hence, it is concluded that once the critical event ``$\gamma _{k,\bar{m}}=1$ and $\gamma_{k,\breve{m}}^{e}=0$ for some $k\ge 0$'' occurs, the eavesdropper's mean estimation error $\{ \mathbb{E} [ \tilde{x}_{k|k,\breve{m}}^{e} ] \} _{k>0}$ diverges under the provided PPM (\ref{eq5})--(\ref{eq7}). 
	$\hfill \blacksquare$
\end{pf}

\begin{remark}
	In \cite{Huang_Auto2021_E}, a novel encryption scheduling method is presented, where the sensor determines whether to encrypt the data before each transmission. 
	This encryption scheduling method is compatible with most encryption techniques. However, it may fail to ensure privacy for systems utilizing the encoding methods , such as those described in \cite{Tsiamis2017_S}, \cite{Tsiamis2018_S}, \cite{Huang2022_P}, and this paper, where the update of the private key relies on the history data. 
	Typically, encryption methods suitable for scheduling, as mentioned in \cite{Huang_Auto2021_E}, impose a significant computational burden. 
	In contrast, the proposed PPM in this paper mitigates computational costs by directly reducing computational complexity to  $\mathcal{O} \left( d_y \right) $, where $d_y$ denotes the dimension of measurements. 
	The process of updating reference data depends on transmission via an acknowledgment channel, where the transmitted message $\gamma _k$ takes values of either 1 or 0, thereby effectively alleviating the communication burden.
\end{remark}

	\section{Simulation Examples}

In this section, we will demonstrate the feasibility and effectiveness of our proposed PPSE method through an Internet-based three-tank system shown as in \cite{He2017_F}. The three-tank system is modelled in the following formula:
\begin{equation}\label{eq46}
	\left\{
	\begin{aligned}
		& x_{k+1}=Ax_k+Bu_k+Dw_k,\\	
		& y_k=Cx_k+Ev_k,
	\end{aligned}
	\right.
\end{equation}
where the input is fixed as $u_k=\left[ \begin{matrix}	3.0\times 10^{-5}&		2.0\times 10^{-5}\\\end{matrix} \right] ^\top$, together with the system parameters and covariance of noises are shown as follows:
\begin{align*}
	& A=\left[ \begin{matrix}	0.9889&		0.0001&		0.0110\\	0.0001&		0.9774&		0.0119\\	0.0110&		0.0119&		0.9770\\ \end{matrix} \right], 
	B=D=\left[ \begin{matrix}	64.5993 &		0.0015\\	0.0015&		64.2236\\	0.3604&		0.3910\\\end{matrix} \right] ,\\
	& C=E=I_{3}, ~Q=10^{-10}I_{3},~R=10^{-4}I_{3}.
\end{align*}
The initial conditions are set as $x_0=\left[ \begin{matrix}	0.3 &0.1&		0.2\\\end{matrix} \right] ^\top$ and $P_0=I_{3}$. 

A two-state MC model is adopted to represent the fading channel, using a probability transition matrix $ M = \left[0.1 ~ 0.9; ~ 0.5 ~ 0.5 \right]$.
Under such a channel, the reception probabilities conditioned on the MC state are $\mathrm{Prob} \left\{ \gamma _k=1|\rho _k=\theta _1 \right\}=0.3$ and $\mathrm{Prob} \left\{ \gamma _k=1|\rho _k=\theta _2 \right\}=0.9$,  same as the interception probabilities for the eavesdropper. The wiretap channel is also characterized by a two-state MC, but with a distinct transition matrix $M_e=\left[  0.2 ~ 0.8; ~ 0.4 ~ 0.6 \right] $. We assess the estimation performance of the legitimate user and the eavesdropper using the mean square error (MSE), calculated via 2000 Monte Carlo simulations.

\begin{figure}
	\begin{center}
		\includegraphics[width=1\linewidth,scale=1]{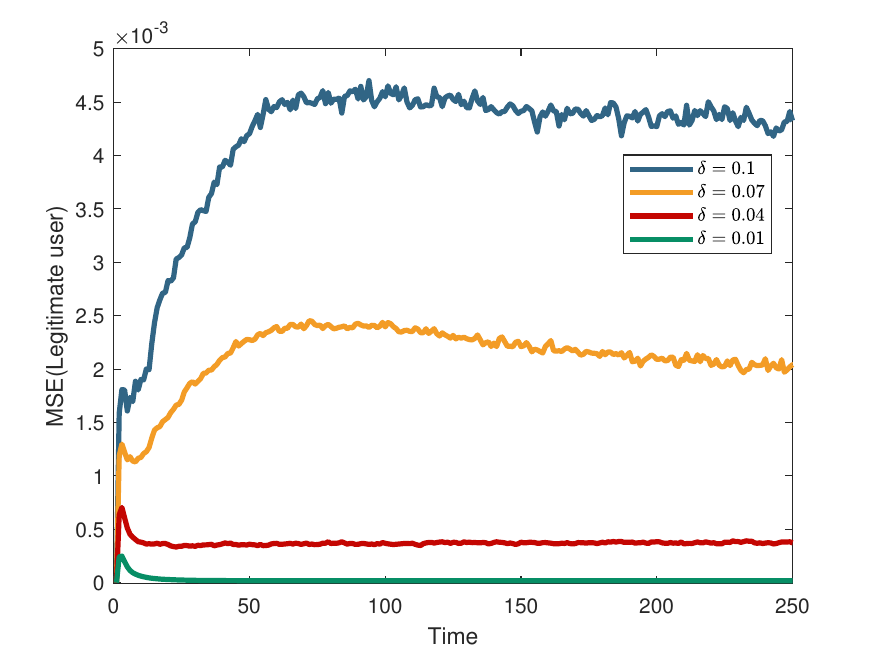}    
		\caption{Trajectories of the legitimate user's MSE under the PPM with different values of $\delta$.}  
		\label{Fig2}                                 
	\end{center}   
\end{figure}

We first analyze the impact of the PPM parameter $\delta$ on the estimation performance of legitimate users. For this purpose, we fix the PPM parameters at $a=2$ and $s=1$, and vary $\delta$ across values of 0.01, 0.04, 0.07, and 0.1, as shown in Fig. \ref{Fig2}.
The results indicate that as $\delta$ increases, the MSE also increases. This behavior arises because, under such conditions, the quantization function $\mathcal{Q} \left( \cdot \right) $ in the PPM becomes coarser, thereby reducing decoding accuracy and negatively affecting estimation performance. 
While a PPM with higher precision improves estimation accuracy, it does so at the cost of increased data length. Consequently, the choice of $\delta$ should be carefully optimized based on the available transmission capacity.

\begin{figure}
	\begin{center}
		\includegraphics[width=1\linewidth,scale=1]{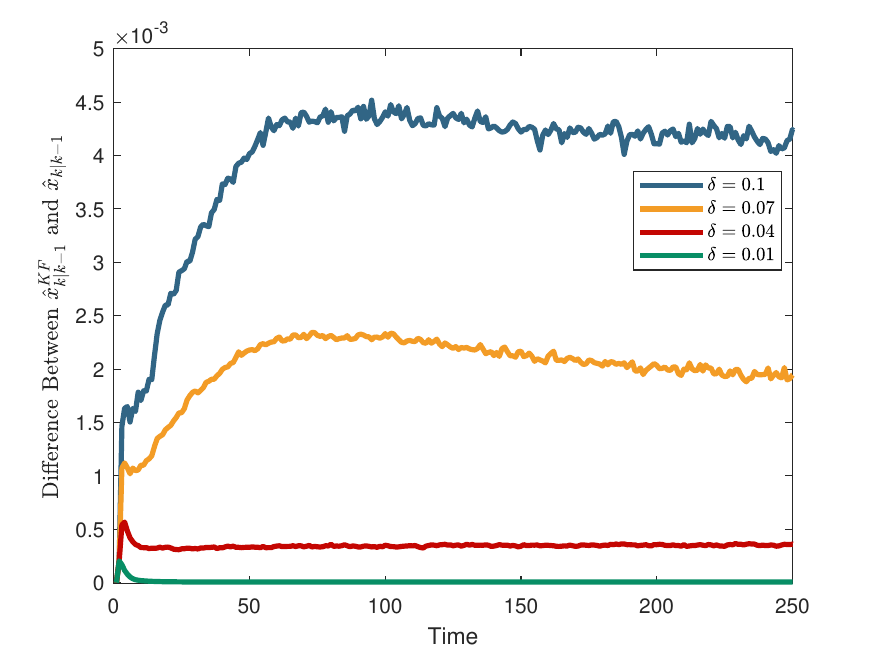}       
		\caption{Trajectories of MSE between state predictions generated by standard KF and PPF.}  
		\label{Fig3}                                 
	\end{center}   
\end{figure}

Fig. \ref{Fig3}  illustrates the differences between the state predictions generated by the standard KF $ \hat{x}_{k|k-1}^{KF} = \mathbb{E} \left[ x_k | y_k \right]  $ and PPF $ \hat{x}_{k|k-1} = \mathbb{E} \left[ x_k | \zeta_k \right]  $ for various values of $ \delta $. 
The results indicate that lower values of $ \delta $ lead to smaller differences, demonstrating that the state predictions of the PPF can closely approximate those generated under linear Gaussian conditions.

\begin{figure}
	\begin{center}
		\includegraphics[width=1\linewidth,scale=1]{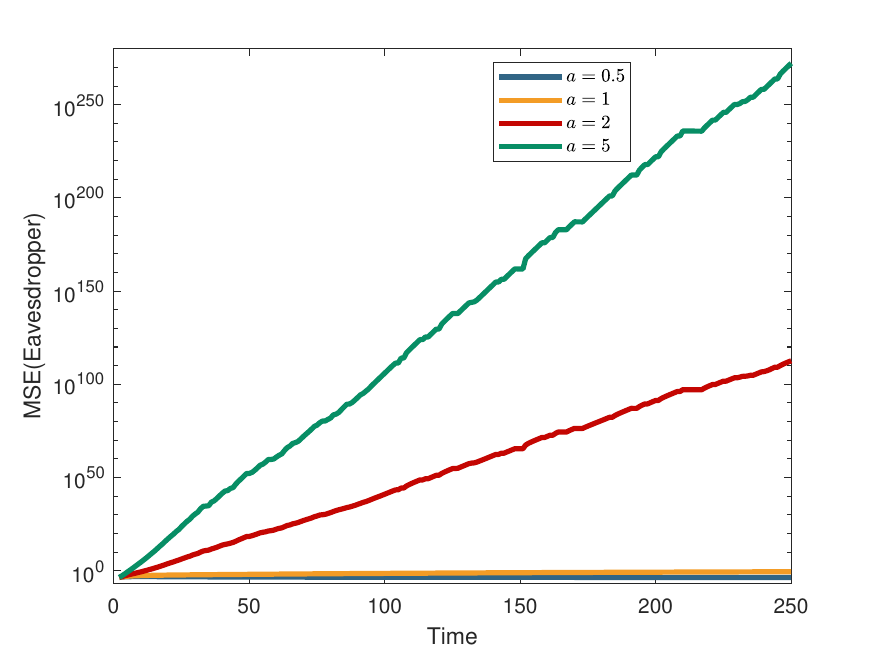}       
		\caption{Trajectories of the eavesdropper's MSE under different encoding parameters $a$.}  
		\label{Fig4}                                 
	\end{center}   
\end{figure} 

\begin{figure}
	\begin{center}
		\includegraphics[width=1\linewidth,scale=1]{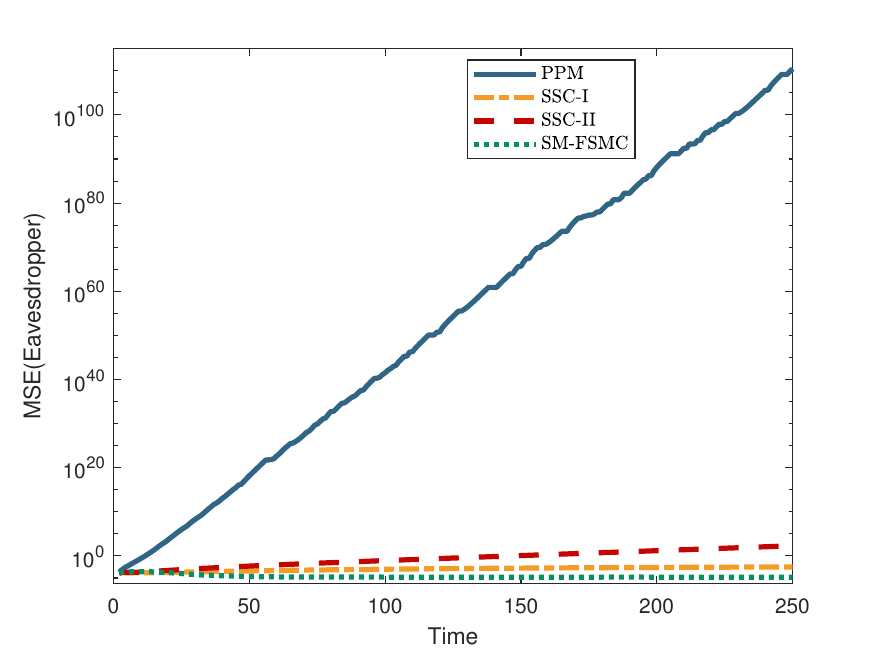}      
		\caption{Trajectories of the eavesdropper's MSEs using different encoding methods in Example 2.}  
		\label{Fig5}                                 
	\end{center}   
\end{figure}

Additionally, we analyze the influence of the scaling parameter $a$ on the divergence rate of the eavesdropper's estimation error, thereby demonstrating how different values of $a$ affect the effectiveness of privacy preservation. 
The PPM parameters are set to be $s=1$ and $\delta=0.01$.
The parameter $a$ is varied across the values $a=0.5,1,2,5$.
The trajectories of the eavesdropper's MSE with these different values of $a$ are depicted in Fig. \ref{Fig4}.
Fig. \ref{Fig4} reveal that when $a > 1$  and a critical event occurs, the eavesdropper's MSE tends to diverge, confirming Theorem 3. 
Furthermore, as the parameter $a$ increases, the eavesdropper's MSE rises rapidly.

Similar to our proposed PPM, the secrecy mechanisms presented in \cite{Tsiamis2020_S}, \cite{Tsiamis2018_S}, and \cite{Impicciatore_CDC2022_S} utilize encoding-decoding as an encryption method and adhere to the principles of control-theoretic perfect secrecy. To demonstrate the effectiveness of the proposed PPM, we compare the eavesdropper's estimation performance derived from the PPM with the performance of other secrecy mechanisms introduced in these works.
The specific encoding structure, referred to as the state-secrecy code (SSC), is introduced in \cite{Tsiamis2020_S} and \cite{Tsiamis2018_S}.
The SSCs are constructed using a weighted reference defined as $z_k=\hat{x}_k-L^{k-t_k}\hat{x}_{t_k}$ and are activated by triggering a critical event. However, the weight matrix $L$  must be separately designed for unstable and stable systems in SSC-I and SSC-II, respectively.
For SSC-I, designed for unstable systems, the weight matrix is set to $L=A$. In SSC-II, designed for stable systems, the weight matrix is given by $L=P_L \left( P_LA^{\top} \right) ^{-1}$, where $P_L$  is the open-loop covariance calculated as $ P_L =A P_L A ^\top + D Q D ^\top$. In this simulation example, the covariance PLPL is $[0.4238 ,~0.1226,~ 0.2361;~0.1226,~0.1536,~0.1156;~ 0.2361,~0.1156,$ $0.1731] \times 10^{-4}$. The PPSE problem over a finite-state Markov channel is explored in \cite{Impicciatore_CDC2022_S}, where a withholding-based strategy is utilized to design the secrecy mechanism, referred to as SM-FSMC. In SM-FSMC, the measurement $z_k$ retains its original value $y_k$ with a certain probability, defined as $\mathrm{Prob}\left\{ z_k=y_k \right\} =\hat{\lambda}$. Otherwise, the measurement is withheld.

The comparison results are illustrated in Fig. \ref{Fig5}, which depicts the trajectories of the eavesdropper's MSE under different methods.
Fig. \ref{Fig5} shows that the bias introduced by SSC-I and SM-FSMC is nearly negligible, indicating that these methods are ineffective. This observation arises because the three-tank system is stable, however, SSC-I and SM-FSMC are only effective for unstable systems. Furthermore, SM-FSMC requires that the legitimate user must have a higher reception probability compared to the eavesdropper.
The red line represents SSC-II, which is specifically designed for stable systems. Consequently, the eavesdropper's MSE diverges under SSC-II. The proposed PPM also leads to divergence in the eavesdropper's MSE.
The frequency of critical events occurring in SSC-II matches that of the PPM. With this consistent frequency, the PPM yields a more rapid increase in the eavesdropper's MSE compared to SSC-II. This indicates a faster divergence rate of the eavesdropper's MSE under PPM. The rate of divergence in the eavesdropper's MSE can be thought of as the level of secrecy. Although both methods facilitate privacy preservation, the PPM has a higher level of secrecy.

\section{Conclusions}

In this paper, we have developed a PPSE framework for cyber-physical systems over MFCs in the presence of an eavesdropper.
By utilizing the acknowledgment channel between the legitimate user and the sensor, we have proposed an expectational lossless PPM based on reference data,  ensuring secrecy with minimal computational costs. 
Our dynamic encoding-decoding strategy exploits packet loss events to degrade eavesdropper's performance and introduces a scalar encoding parameter that amplifies the eavesdropper's estimation error without degrading the legitimate user's performance.
Specifically, we have designed a recursive filter tailored to the PPM, balancing computational efficiency and estimation accuracy for the legitimate user.   
We have analyzed the exponential boundedness of the legitimate user's estimation error covariance, illustrating the impact of the stochastic properties of MFC and the distortion rate of encoding on the boundedness.
Additionally, we have examined the divergence of eavesdropper's estimation error.
Finally, we have provided a simulation example of a three-tank system to demonstrate the effectiveness and feasibility of our PPSE approach, showing that co-designing the PPSE can ensure secrecy in both stable and unstable systems. 
% Comparative studies with existing methods reveal superior privacy performance, owing to the amplification effect of the scaling parameter.

\bibliographystyle{model5-names}         
\bibliography{refs}

\end{document}